\documentclass[manuscript]{aastex}
\usepackage{enumerate}
\usepackage{xcolor}
\shorttitle{Restoring process of a sunspot penumbra}
\shortauthors{Romano et al.}

\begin{document}

%% LaTeX will automatically break titles if they run longer than
%% one line. However, you may use \\ to force a line break if
%% you desire.

\title{Restoring process of sunspot penumbra}

%% Use \author, \affil, and the \and command to format
%% author and affiliation information.
%% Note that \email has replaced the old \authoremail command
%% from AASTeX v4.0. You can use \email to mark an email address
%% anywhere in the paper, not just in the front matter.
%% As in the title, use \\ to force line breaks.

\author{P. Romano\altaffilmark{1}, M. Murabito\altaffilmark{2}, S.L. Guglielmino\altaffilmark{3}, F. Zuccarello\altaffilmark{3} and M. Falco\altaffilmark{1}}%\affil{Astronomy Department, University of California,
%    Berkeley, CA 94720}

%\author{C. D. Biemesderfer\altaffilmark{4,5}}
%\affil{National Optical Astronomy Observatories, Tucson, AZ 85719}
\email{paolo.romano@inaf.it}

%\and

%\author{R. J. Hanisch\altaffilmark{5}}
%\affil{Space Telescope Science Institute, Baltimore, MD 21218}

%% Notice that each of these authors has alternate affiliations, which
%% are identified by the \altaffilmark after each name.  Specify alternate
%% affiliation information with \altaffiltext, with one command per each
%% affiliation.

\altaffiltext{1}{INAF - Osservatorio Astrofisico di Catania,
              Via S.~Sofia 78, 95123 Catania, Italy.}
\altaffiltext{2}{INAF - Osservatorio Astronomico di Roma, 
							Via Frascati 33, I-00040 Monte Porzio Catone, Italy.}							
\altaffiltext{3}{Dipartimento di Fisica e Astronomia ``Ettore Majorana'' - Sezione Astrofisica, Universit\`{a} di Catania,
			 Via S.~Sofia 78, 95123 Catania, Italy}
		 			 
%\altaffiltext{3}{present address: Center for Astrophysics,
%    60 Garden Street, Cambridge, MA 02138}
%\altaffiltext{4}{Visiting Programmer, Space Telescope Science Institute}
%\altaffiltext{5}{Patron, Alonso's Bar and Grill}

%% Mark off your abstract in the ``abstract'' environment. In the manuscript
%% style, abstract will output a Received/Accepted line after the
%% title and affiliation information. No date will appear since the author
%% does not have this information. The dates will be filled in by the
%% editorial office after submission.

\begin{abstract}
We describe the disappearance of a sector of sunspot penumbra and its restoring process observed in the preceding sunspot of active region NOAA 12348. The evolution of the magnetic field and the plasma flows support the idea that the penumbra forms due to a change of inclination of the magnetic field of the canopy. Moving magnetic features have been observed during the disintegration phase of that sector of sunspot penumbra. During the restoring phase we have not observed any magnetic flux emergence around the sunspot. The restoring process of the penumbra sector completed in about 72 hours and it was accompanied by the transition from the counter-Evershed flow to the classical Evershed flow. The inversion of photospheric spectropolarimetric measurements taken by IBIS allowed us to reconstruct how the uncombed configuration of the magnetic field forms during the new settlement of the penumbra, i.e., the vertical component of the magnetic field seems to be progressively replaced by some horizontal field lines, corresponding to the intra-spines. 
\end{abstract}

%% Keywords should appear after the \end{abstract} command. The uncommented
%% example has been keyed in ApJ style. See the instructions to authors
%% for the journal to which you are submitting your paper to determine
%% what keyword punctuation is appropriate.


%% From the front matter, we move on to the body of the paper.
%% In the first two sections, notice the use of the natbib \citep
%% and \citet commands to identify citations.  The citations are
%% tied to the reference list via symbolic KEYs. The KEY corresponds
%% to the KEY in the \bibitem in the reference list below. We have
%% chosen the first three characters of the first author's name plus
%% the last two numeral of the year of publication as our KEY for
%% each reference.

%% Authors who wish to have the most important objects in their paper
%% linked in the electronic edition to a data center may do so by tagging
%% their objects with \objectname{} or \object{}.  Each macro takes the
%% object name as its required argument. The optional, square-bracket 
%% argument should be used in cases where the data center identification
%% differs from what is to be printed in the paper.  The text appearing 
%% in curly braces is what will appear in print in the published paper. 
%% If the object name is recognized by the data centers, it will be linked
%% in the electronic edition to the object data available at the data centers  
%%
%% Note that for sources with brackets in their names, e.g. [WEG2004] 14h-090,
%% the brackets must be escaped with backslashes when used in the first
%% square-bracket argument, for instance, \object[\[WEG2004\] 14h-090]{90}).
%%  Otherwise, LaTeX will issue an error. 
\section{Introduction}

Sunspot formation is the main manifestation of the magnetic flux emergence from the convection zone into the solar atmosphere. Among the different features characterizing sunspots, the penumbra with its properties, formation, and evolution is of particular interest for the scientific community, because there are still several issues to be solved (\citealp{Rolf:09,Borrero:11}; see also Section~7.1 in \citealp{Hinode10}). The most important one is its bolometric brightness (about 75\% of the quiet photosphere on average) which requires large vertical velocities ($\approx$ 1-2~km~s$^{-1}$) that are not usually observed \citep{Lan05}, although they are necessary to justify the heat transport from the sub-photospheric zone into the photosphere. 

The main features forming the penumbra are thin, radially elongated filaments that are central for heat transport, although the interpretation of the observations of plasma flows along the penumbral filaments and the corresponding magnetic field inclination are not trivial. Two main models have been proposed to account for the filamentary structure of penumbrae: the embedded flux tube model \citep{Sol93, Sch98a, Sch98b, Bel03, Bel04} and the field-free gap model \citep{Sch06, Spr06}. The former suggests that the part of the flux tube that is in contact with the underlying hotter quiet Sun is heated by radiation, expands, gets less dense than the surroundings, and rises as a result of buoyancy. In this case, rising magnetic flux tubes embedded in more vertical background magnetic fields carry the Evershed flow, i.e., the plasma is accelerated outwards from the umbra-penumbra boundary to the outer edges of the sunspots. By this gas pressure gradient, the flow should be accelerated from 3~km~s$^{-1}$ at the inner footpoint up to 14~km~s$^{-1}$ near the outer edge of the penumbra. The field-free gap model interprets the penumbra as consisting of bright filaments in the shape of elevated bright structures on a dark background. A penumbral filament is the surface manifestation of a field-free gap below the observed surface, communicating directly with the surrounding convection zone. 

%Recent MHD simulations have been successful in reproducing many aspects of the fine structure of penumbrae \citep{Rem11, Rem12, Rem15}. They show that the magnetic field structures and their association with horizontal flows close to the solar surface are compatible with the embedded flux model, although the presence of azimuthal convection throughout the penumbra is closer to the idea of \citet{Spr06}.

In recent years, MHD numerical simulations have been more and more successful in reproducing many aspects of the penumbral fine structure (\citealp{Heinemann:07, Brummell:08, Scharmer:08, Rempel:09a, Rempel:09b, Rem11, Rem12, Rem15, Chen:17}; see also the review by \citealp{Rempel:11}). They are partially in agreement with the embedded flux model for the description of the magnetic field configuration of the penumbra and the horizontal flows along the main axes of the filaments. However, they also show the presence of the azimuthal convection that is compatible with the field-free gap model. Therefore, further studies are necessary to discriminate between the two models.

An important progress in the knowledge of the penumbra has been obtained by \citet{Tiw13}, who provided a global picture of the penumbra fine structure using Hinode observations. They found that the brightness of a filament is largest at its head, i.e., the inner footpoint, and decreases towards its tail, i.e., the outer edge; the magnetic inclination of penumbral filaments looks like a strongly flattened $\bigcap$-loop; the temperature of the environment in which the filaments are embedded increases systematically from the inner to the outer penumbra; strong upflows are observed in the inner parts of penumbrae and strong downflows in their outer part, the upflows extend along the whole axis of each filament and the horizontal velocity has also an azimuthal component.

There is still a debate about two main proposed scenarios for penumbra formation, as described in the following. First, \citet{Lek98} suggested that emerging horizontal field lines form the penumbra because they become trapped by the overlying magnetic field rather than continuing to rise. This idea comes from the detection of a magnetic flux threshold: it seems that above $1.5 \times 10^{20}$~Mx a pore reaches enough total magnetic flux to form the penumbra around it. A second scenario comes from observations of the overlying layers of the solar atmosphere. \citet{Shi12, Rom13, Rom14} observed at chromospheric level the formation of an annular zone around the corresponding location of a pore in photosphere, before the appearance of its penumbra. These observations suggest that the penumbra formation is a top-down process: the magnetic field, already emerged from the convection zone and forming a magnetic canopy above the pore, changes its inclination, bending down to the photosphere and creating the physical conditions for a different heating transport mechanism, which is responsible for the penumbra properties \citep{Mur16}. Actually, the presence of canopy fields with a more horizontal configuration with respect to the solar photosphere has been proposed to lead to the formation of penumbral-like structures as well \citep[e.g.,][]{Zuccarello:14,Guglielmino:17,Guglielmino:19}. 

In this context, the transition of the Evershed flow direction at photospheric level before and after the penumbral filament settlement appears to be a signature of penumbra formation  \citep{Mur16}. It seems that the Evershed flow is usually inwards to the proto-spot center in the photosphere before the penumbra formation (counter-Evershed flow) and becomes outwards after the penumbra formation is completed (classical Evershed flow). This may be due to the settlment of penumbral-like  connection between small flux concentrations (e.g., pores) adjacent to the main sunspot during the first phases of penumbra formation \citep{SiuTapia:17}. Nonetheless, this photospheric counter-Evershed flow should not be confused with the inverse-Evershed flow that is usually observed in the chromosphere cospatially to the photospheric Evershed flow (\citealp{Maltby:75}; see also \citealp{Beck:20}). 

It has been suggested that the emergence of new magnetic flux from the convection zone is not compatible with the penumbra formation. In fact, \citet{Sch10} found that the activity and dynamics due to the ongoing flux emergence prevent that the penumbra can settle down. Nevertheless, it is noteworthy that \citet{Mur17, Mur18} observed also the settlement of the penumbra between the main opposite magnetic polarities where new magnetic flux was still emerging.

The decay phase is also useful to clarify the physics which describes the coupling of the plasma and the magnetic field in stable penumbral filaments.
\citet{Ben18} observed that during the decay phase of a sunspot, contrarily to what is seen in stable sunspots, the inner penumbra boundary does not match with a constant value of the vertical magnetic field, i.e. the invariant vertical component of the magnetic field of the umbra-penumbra boundary found by \citet{Jur17a}. In fact, the umbra does not have a sufficiently strong vertical component of the magnetic field, so that the penumbra becomes unstable and prone to be destroyed by convection or magnetic diffusion. 

Usually the presence of light bridges (LBs) is an indication of sunspot decay phase and incoming sunspot fragmentation \citep[e.g.,][and references therein]{Fal16, Fel16}. LBs rapidly intrude from the leading edge of penumbral filaments into the umbra and re-establish granular motions within the spot. Recently, using near-infrared spectropolarimetric data taken by the GREGOR solar telescope, \citet{Ver18} observed that a LB initiated the transformation of penumbral filaments into elongated, dark umbral cores and induced a variation of the magnetic field inclination from the horizontal direction in the penumbra into a more vertical field, typical of sunspot umbra. The penumbral sector involved by this decay process, which was triggered by the interaction between emerging and already established flux systems, was characterized by weak Evershed flow of about 0.1~km~s$^{-1}$ and a poor presence of moat flow, as observed in the \ion{Si}{1} maps at 1082.7 nm.

In this paper, we consider high-resolution observations obtained by the Interferometirc Bidimensional Spectroscopic Instrument \citep[IBIS;][]{Cav06} which operated at the NSO/Dunn Solar Telescope (DST), acquired during the decay and subsequent re-formation phases of a sector of penumbra in the preceding sunspot of Active Region (AR) NOAA 12348.  The reconstruction of the magnetic field properties and the measurement of the plasma flow by means of these spectropolarimetric data allowed us to find new indications about the mechanisms that determine at photospheric level the disappearance and  the new settlement of the penumbral filaments, and to shed light on the properties of those penumbral filaments. We also used data acquired by the Helioseismic and Magnetic Imager instrument \citep[HMI;][]{Sche12} onboard the Solar Dynamic Observatory \citep[SDO;][]{Pes12} to study the evolution of the sunspot with continuity. In the next Section, we provide a detailed description of the data used and the main methods applied to get our results, which are reported in Section~3. The last Section is dedicated to the discussion of the results and to draw our conclusions.

\section{Observations and Data analysis}

We observed the disappearance of a sector of the penumbra in the preceding sunspot of AR NOAA 12348 and, later, its restoring process using data acquired by HMI/SDO and by DST/IBIS. 

Space-Weather Active Region Patches \citep[SHARPs;][]{Bob14} data taken by HMI from 18 May, 2015, at 14:00~UT to 21 May, 2015, at 13:48~UT allowed us to describe the evolution of the sunspot exploiting data acquired along the \ion{Fe}{1} line at 617.3~nm. We used the continuum images, the vector magnetic field data and the Dopplergrams, with a time cadence of 12 min and a pixel size of 0\farcs5. This information have been obtained by the inversion of the Stokes data using the Very Fast Inversion of the Stokes Vector \citep[VFISV;][]{Bor11} code, which assumes a Milne-Eddington model of the solar atmosphere. We calibrated the Dopplergrams, assuming that the quiet Sun area comprised in the observed region should exhibit an average line-of-sight (LOS) velocity value of about $-70$~m~s$^{-1}$, corresponding to the convective blueshift tabulated by \citet{Balthasar} for the \ion{Fe}{1} 617.3~nm line. 
%at $\mu=0.922$.}
%assuming that there is no plasma motion in the umbra \citep{Rim94}. For this reason, we calculated the mean line-of-sight (LOS) velocity in the umbra, defined by the normalized continuum intensity, $I_{c} < 0.5$, and subtracted it from the LOS velocity in each pixel of the FOV. {The average value of velocity in umbra used for the normalization of all Dopplergrams was 0.2~km~s$^{-1}$.}

The high-resolution data, with a pixel scale of 0\farcs095, taken by IBIS on 2015, May 18 and 20, were used to investigate in detail the magnetic field configuration at the photospheric level and the morphology at the chromospheric level. We acquired scans along two photospheric lines in full polarimetric mode (\ion{Fe}{1} 630.25 nm and \ion{Fe}{1} 617.3~nm lines) and two chromospheric lines without polarimetric measurements (\ion{Ca}{2} 854.2~nm and H$\alpha$ 656.28~nm lines). The cadence of each scan was about 67~s. The number of scans and the observation time intervals are reported in Table \ref{table1}. All of these lines were sampled with a spectral FWHM of 2~pm, an average spectral step of 2~pm, and an integration time of 60~ms. The sampling was of 30 spectral points along the \ion{Fe}{1} 630.25~nm line, 24 spectral points along the \ion{Fe}{1} 617.3~nm line, 25 spectral points along the \ion{Ca}{2} 854.2~nm line, and 17 spectral points along the H$\alpha$ 656.28~nm line \citep[see][for further details]{Rom17}. The field of view (FOV) of IBIS was $500 \times 1000$ pixels, but we considered often in our analysis only a sub-FOV of $370 \times 370$ pixels centered on the preceding sunspot of the AR. We also acquired broadband images at 633.3~nm simultaneously with the spectral frames, imaging the same FOV with the same exposure time. The spectra have been normalized to the quiet Sun continuum, $I_{c}$. We restored images using the Multi-frame Blind Deconvolution \citep[MFBD;][]{Lof02} technique to reduce the seeing degradation, achiving a spatial resolution of about 0\farcs25 at 630.25~nm. 

\begin{table}
\caption{Scans of \ion{Fe}{1} 630.25~nm, \ion{Fe}{1} 617.3~nm lines, \ion{Ca}{2} 854.2~nm and H$\alpha$ 656.28~nm lines taken by IBIS.\label{table1}}
\centering
\begin{tabular}{c c c c}
\tableline
\tableline
Date &	Start (UT) & End (UT) & Number of scans\\ 
\hline
18-5-2015 & 14:42 & 15:14 & 30\\
18-5-2015 & 15:46 & 16:06 & 20\\
20-5-2015 & 13:28 & 13:39 & 10\\
\tableline
\end{tabular}
\end{table}

We used the SIR code \citep{Rui92} to determine the LOS plasma velocity, magnetic field strength, inclination, and azimuth angles, by performing a single-component inversion of the Stokes profiles of the Fe I 617.3 nm line. We considered three initial guess models for the atmosphere on the base of different thresholds in the continuum intensity Ic, identifying different physical conditions, following \citet{Mur16}. For $I_{c} > 0.9$, corresponding to the quiet Sun, we used as an initial guess the temperature stratification of the Harvard-Smithsonian Reference Atmosphere \citep[HSRA;][]{Gin71} and a value of 0.1~km~s$^{-1}$ for the LOS velocity. For $0.7 < I_{c} < 0.9$ we considered as initial guess a penumbra model where the temperature (T) and the electron pressure changed according to the penumbral stratification indicated by \cite{Del94}. In this case we used an initial value for the magnetic field strength and for LOS velocity of 1000~G and +1~km~s$^{-1}$, respectively. For $I_{c} < 0.7$ we changed the initial T and electron pressure using the values provided by \citet{Col94} (an umbral model for a small spot), with a value of 2000~G for the magnetic field strength. All the physical quantities were assumed to be height independent, while the temperature stratification of each component was modified with three nodes. We modeled the stray-light contamination by averaging over all Stokes I spectra in the 64 pixels characterized by the lowest polarization degree. The spectral point-spread function of IBIS \citep{Rea08} was used to take into account the finite spectral resolution of the instrument.

The 180$^{\circ}$ ambiguity for the azimuthal component was disambiguated and the components of the vector magnetic field transformed into the local solar frame by using the Non-Potential Field Calculation code \citep{Geo05}.

\section{Results}

\subsection{Penumbra evolution as observed by HMI}

During its passage across the solar disc AR NOAA 12348 appeared as formed by a single sunspot of positive polarity. Only some pores with negative polarity were visible around the main spot in some phases of its lifetime. The peculiarity of this target is given by the fact that during its passage across the central meridian the sunspot showed the formation of a LB and the subsequent disruption of part of its umbra and penumbra, but after a few days the penumbra reformed in the same previous location. 

\begin{figure}
\begin{center}
\includegraphics[trim=5 80 280 430, clip, scale=0.6]{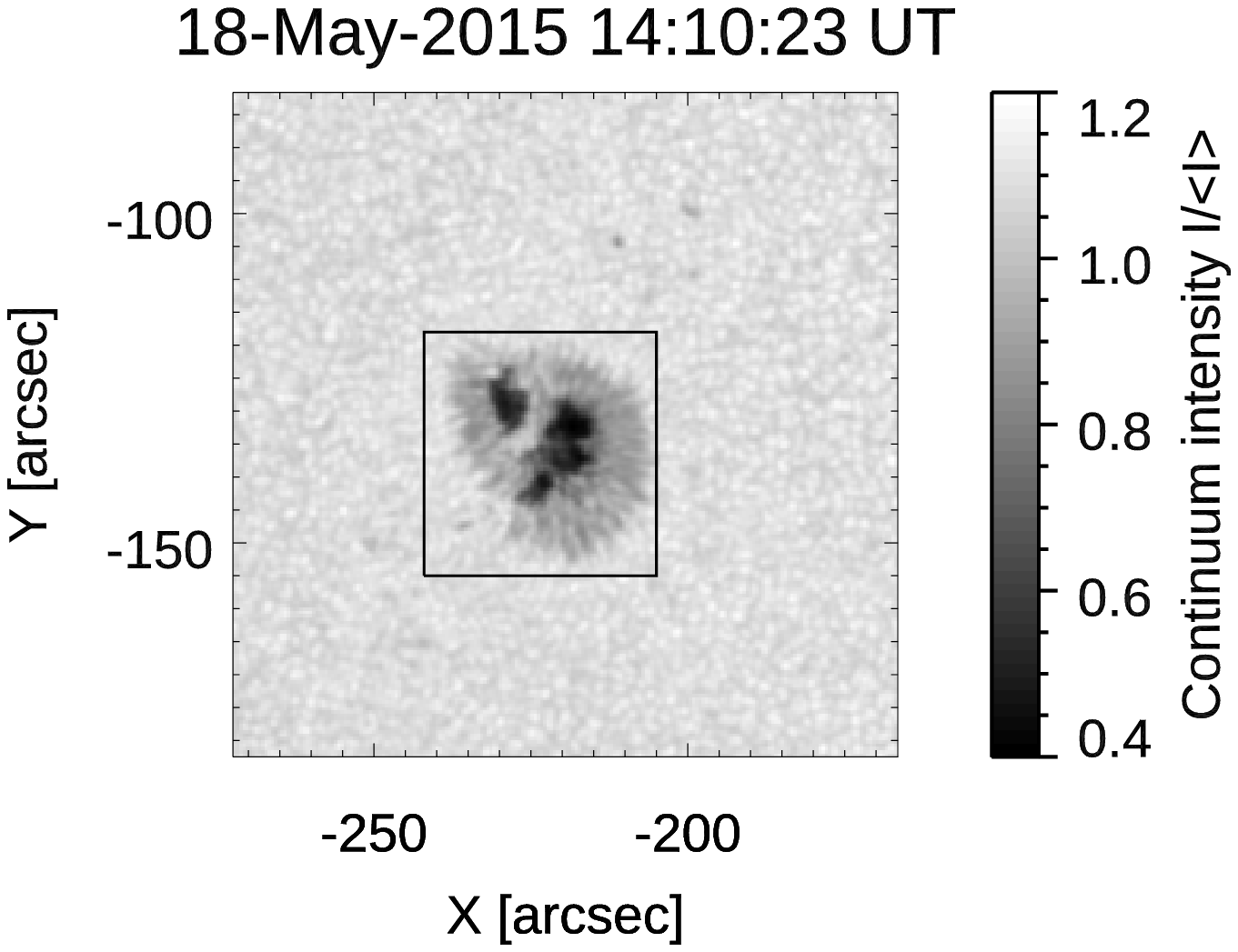}
\includegraphics[trim=35 80 100 430, clip, scale=0.6]{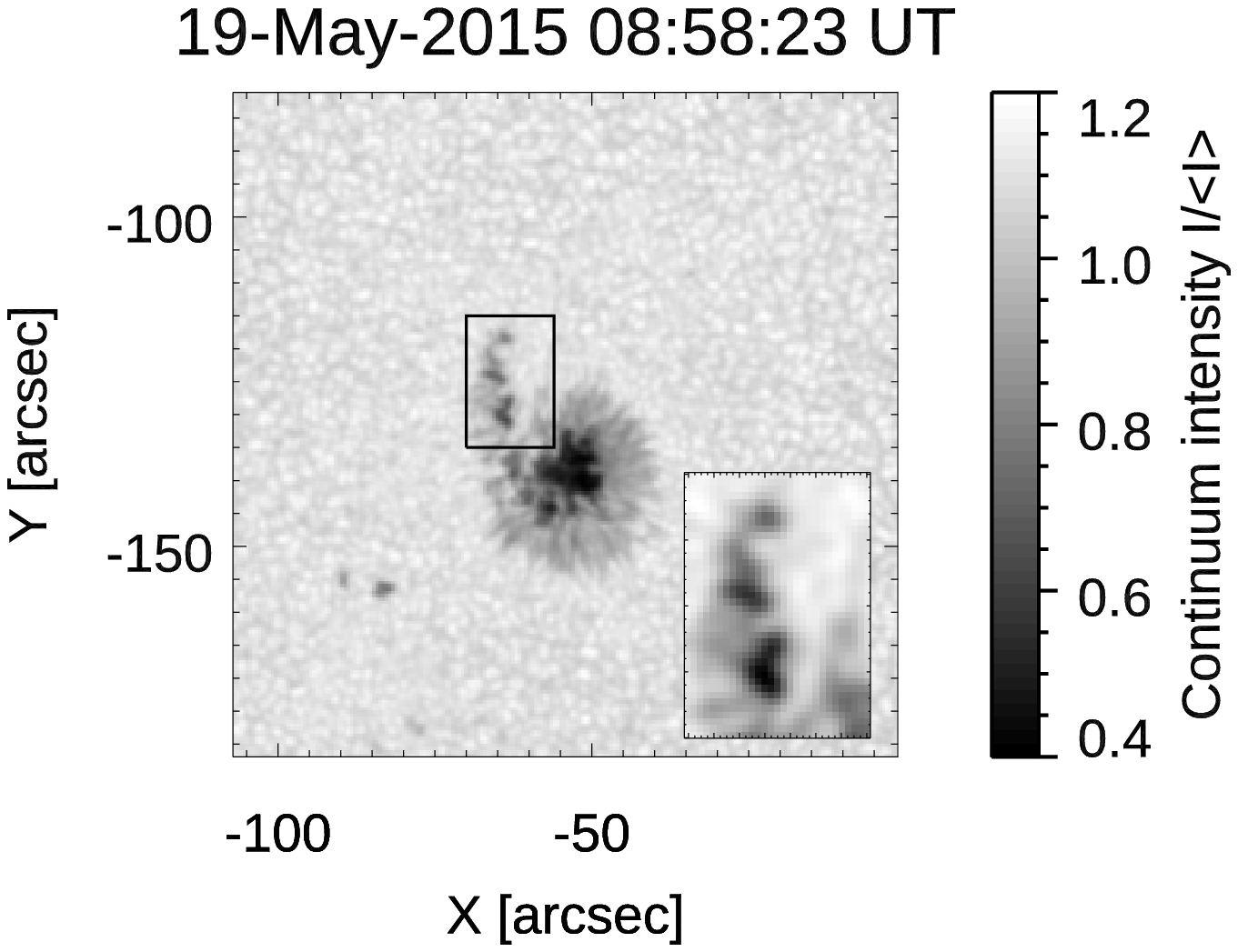}\\
\includegraphics[trim=5 80 280 430, clip, scale=0.6]{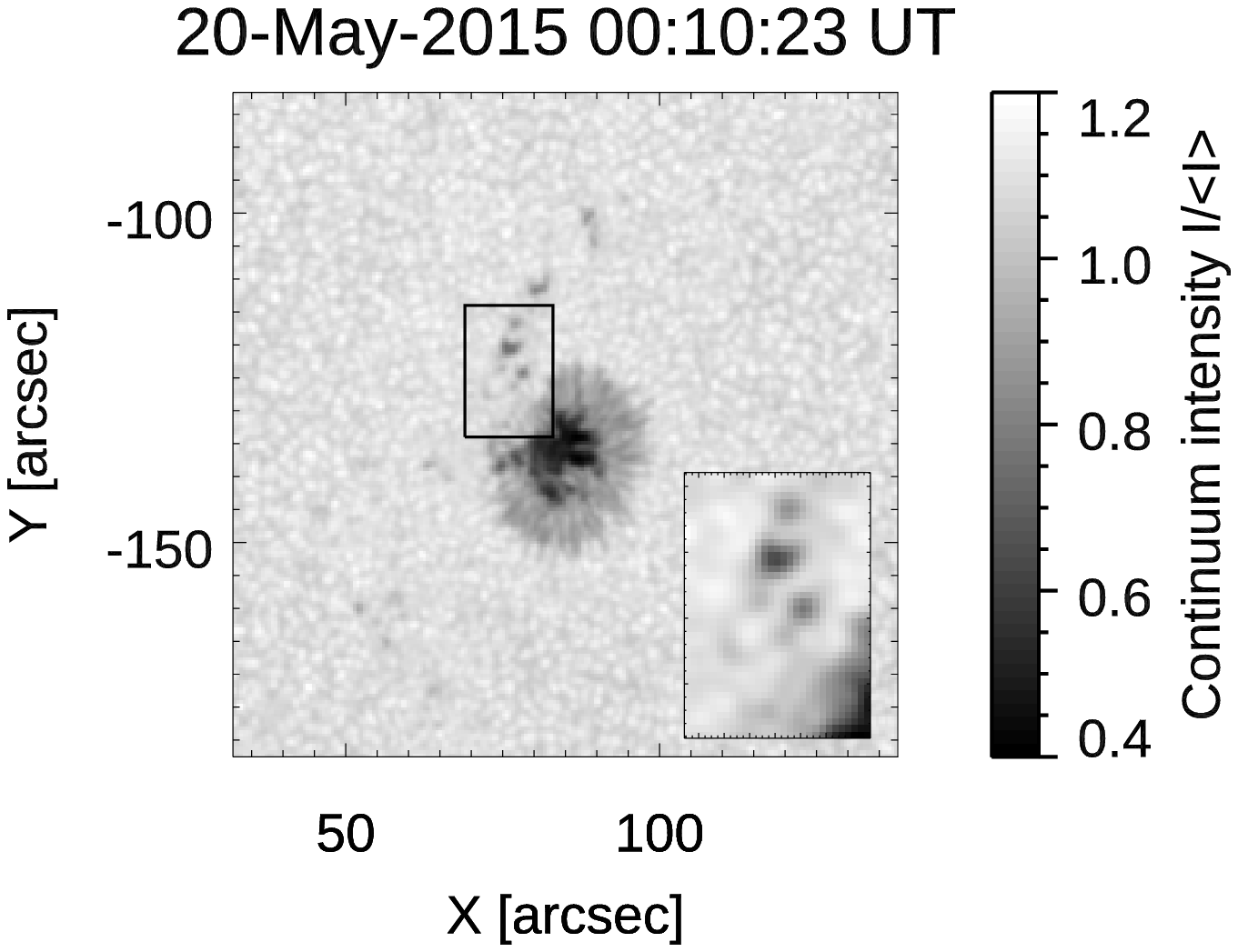}
\includegraphics[trim=35 80 100 430, clip, scale=0.6]{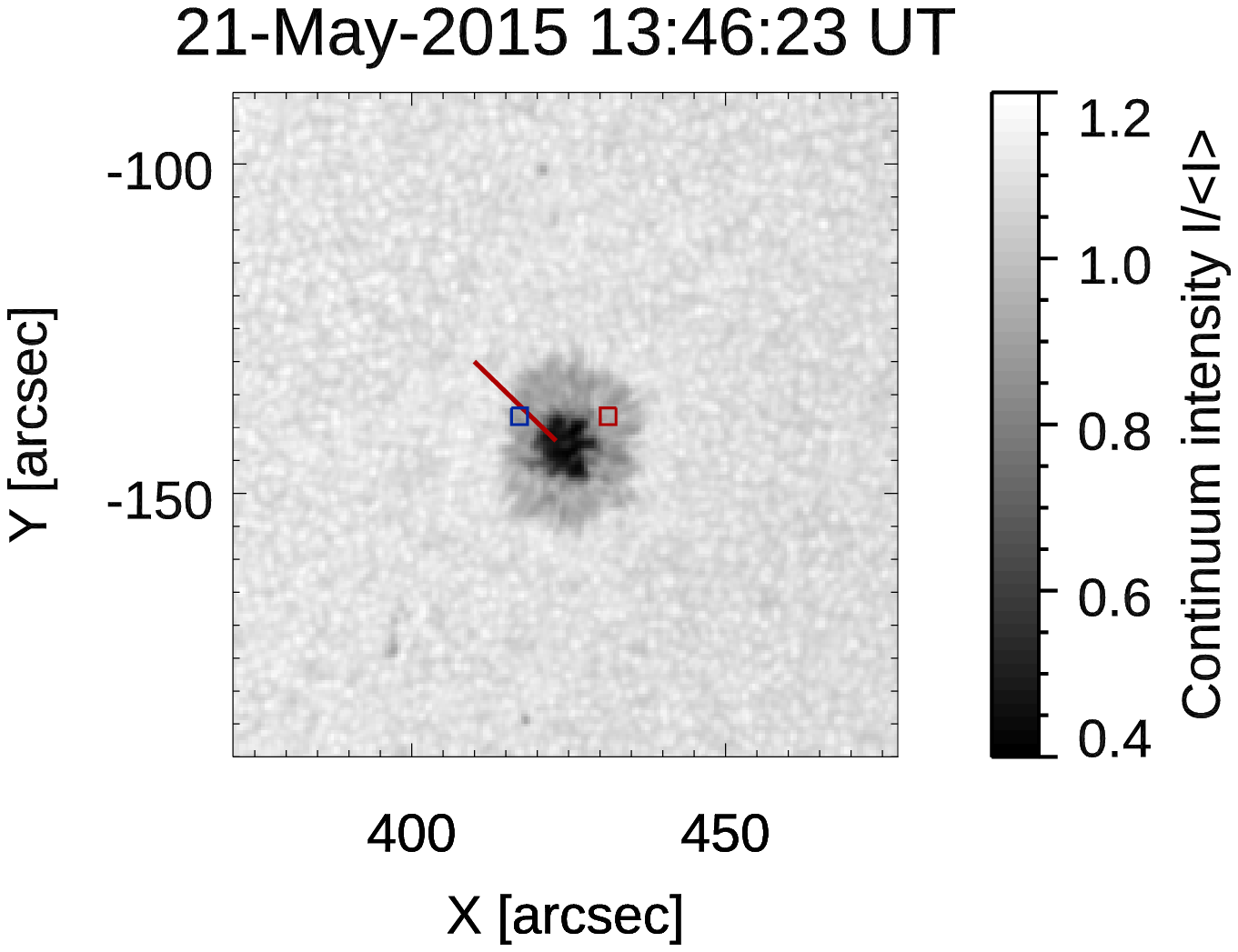}
\caption{ SHARP continuum images of AR NOAA 12348 acquired by HMI/SDO. The box in the top left panel indicates the IBIS FOV reported in Figures~\ref{fig5}, \ref{fig7} and~\ref{fig8}. In the bottom right panel the red and blue boxes indicates the areas where the evolution of the parameters reported in Figure~\ref{fig10} has been measured, while the red segment indicates the radial cut where the continuum intensity variation reported in the top left panel of Figure~\ref{fig9} has been computed. More details of the continuum evolution can be seen in the accompanying online movie ($Continuum\_AR12348.wmv$; see main text). \label{fig1}} 
\end{center}
\end{figure}

On 2015, May 18, the HMI continuum image shows the umbra of the sunspot divided in two portions, with the LB crossing the sunspot along the direction from South-East to North-West (top left panel of Figure~\ref{fig1}). Then, the smallest portion of the sunspot was disrupted by the fragmentation of the umbra into several parts and the disappearance of the penumbral filaments (top right panel of Figure~\ref{fig1}). On May 20 the sunspot appeared missing of some penumbral sectors where the granulation took over (bottom left panel of Figure~\ref{fig1}). Surprisingly, the penumbra started to reform counterclockwise from azimuth angles $120^{\circ}$ to about $210^{\circ}$\footnote{Azimuth angles are measured counterclockwise from $0^{\circ}$ pointing to solar West. On} May 21 the whole penumbra was observed again. The restoring process of the penumbra completed on May 21 at around 14:00~UT (bottom right panel of Figure \ref{fig1}), i.e., after 72 hours from the beginning of the penumbra disruption. During the penumbra restoring process the sunspot was characterized by a proper motion, i.e., it moved in the South-East direction. We also clearly see that the umbra and the whole sunspot reduced their size.

\begin{figure}
\begin{center}
\includegraphics[trim=5 80 280 430, clip, scale=0.6]{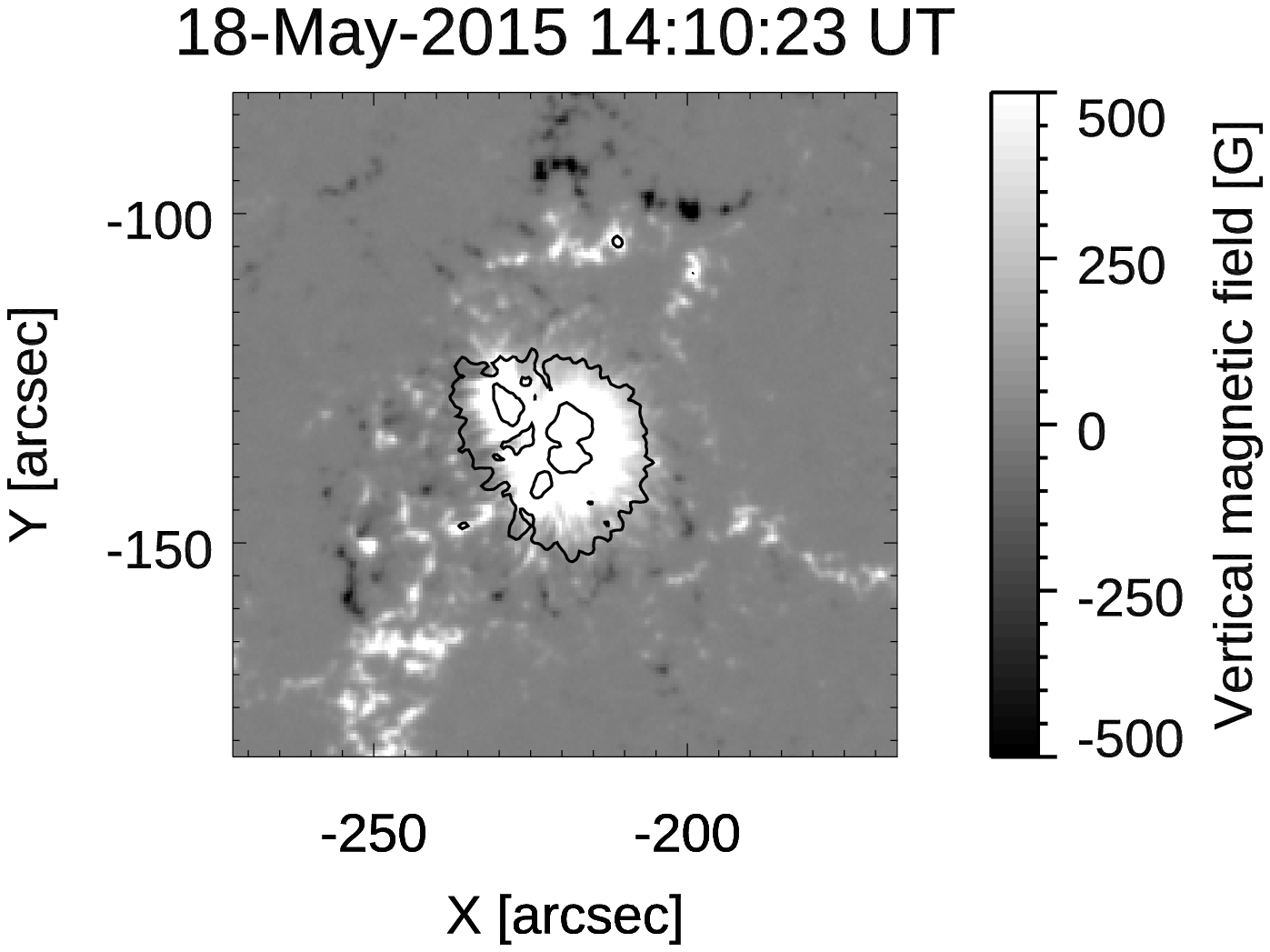}
\includegraphics[trim=35 80 100 430, clip, scale=0.6]{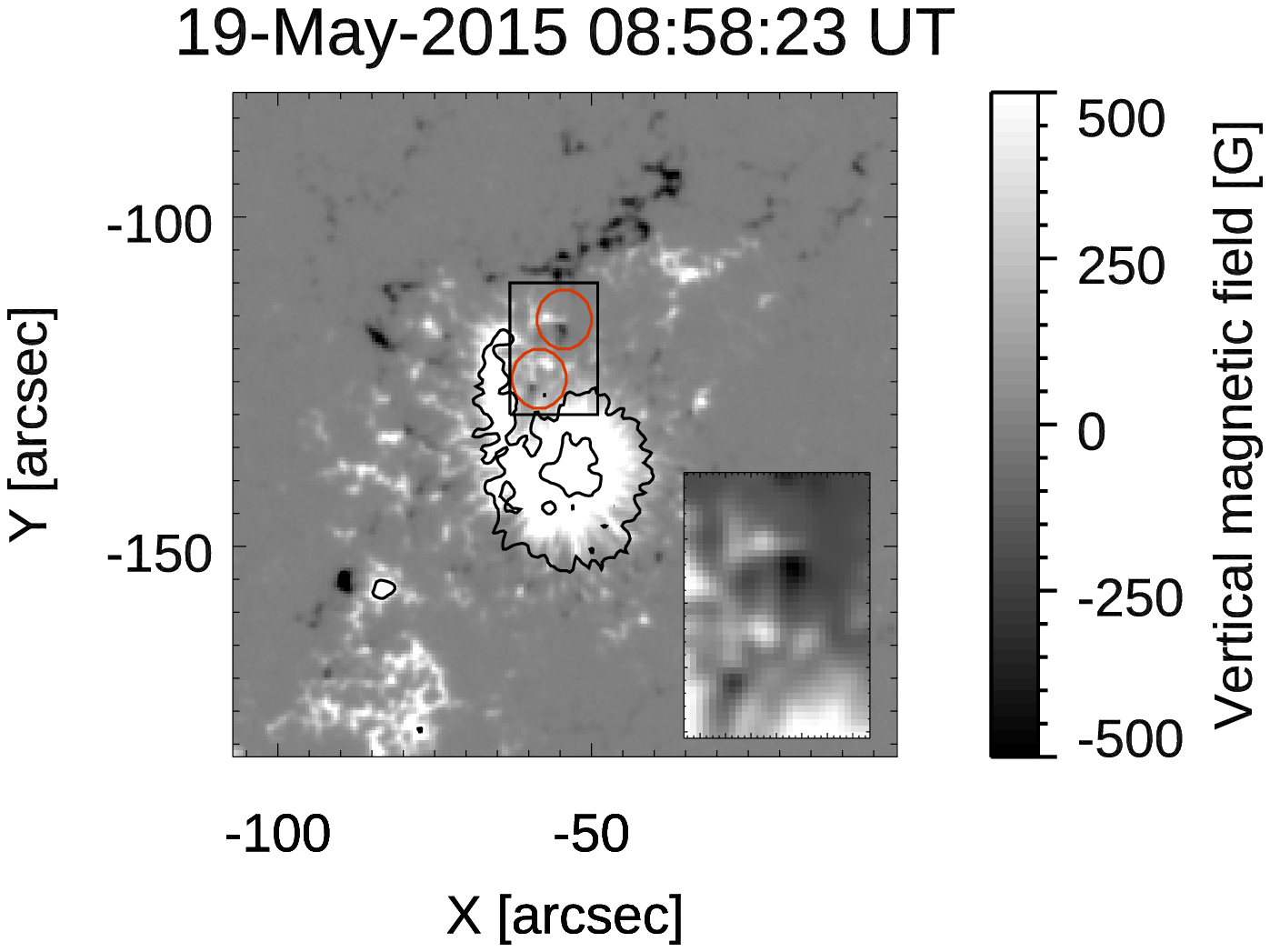}\\
\includegraphics[trim=5 80 280 430, clip, scale=0.6]{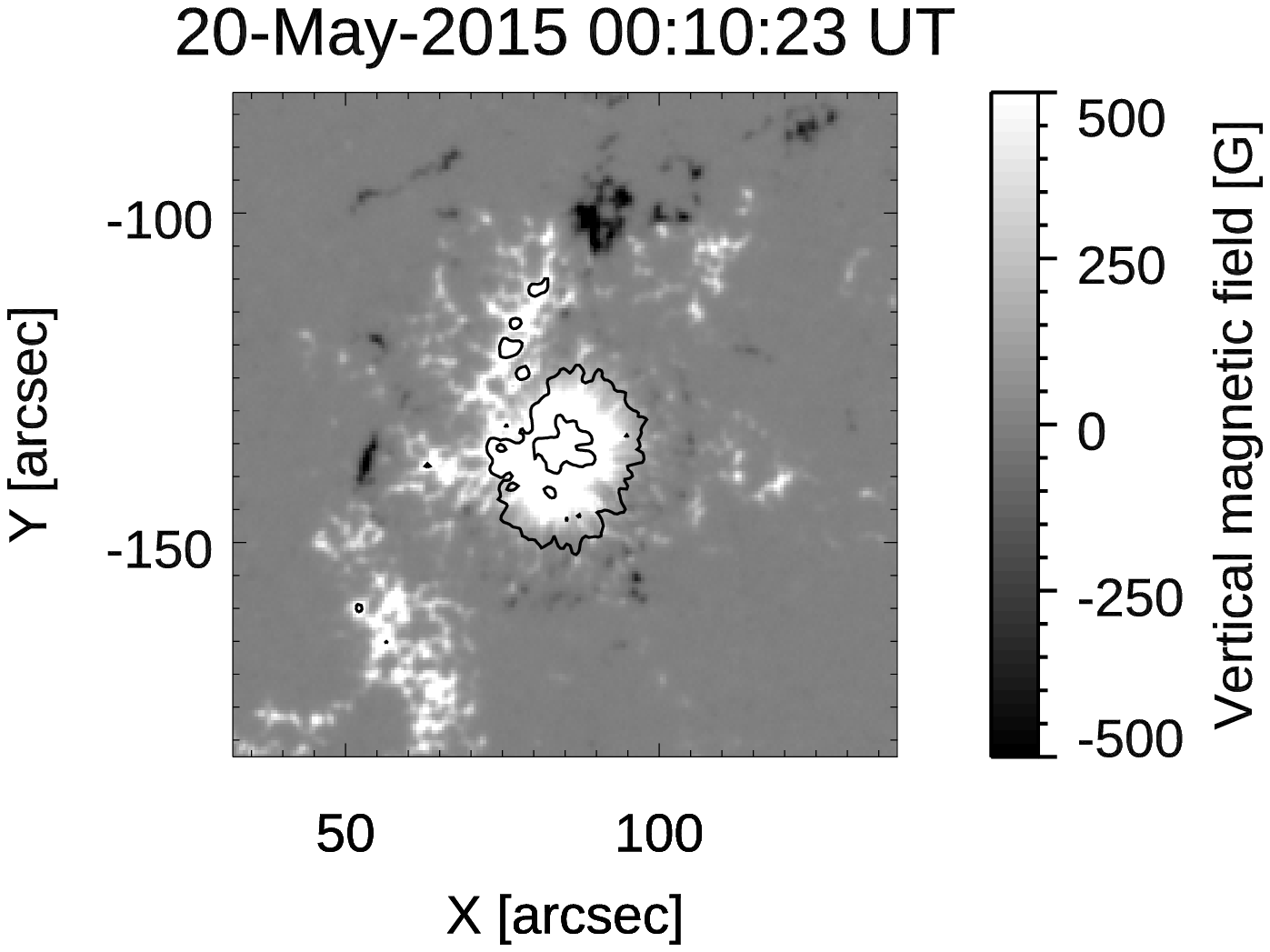}
\includegraphics[trim=35 80 100 430, clip, scale=0.6]{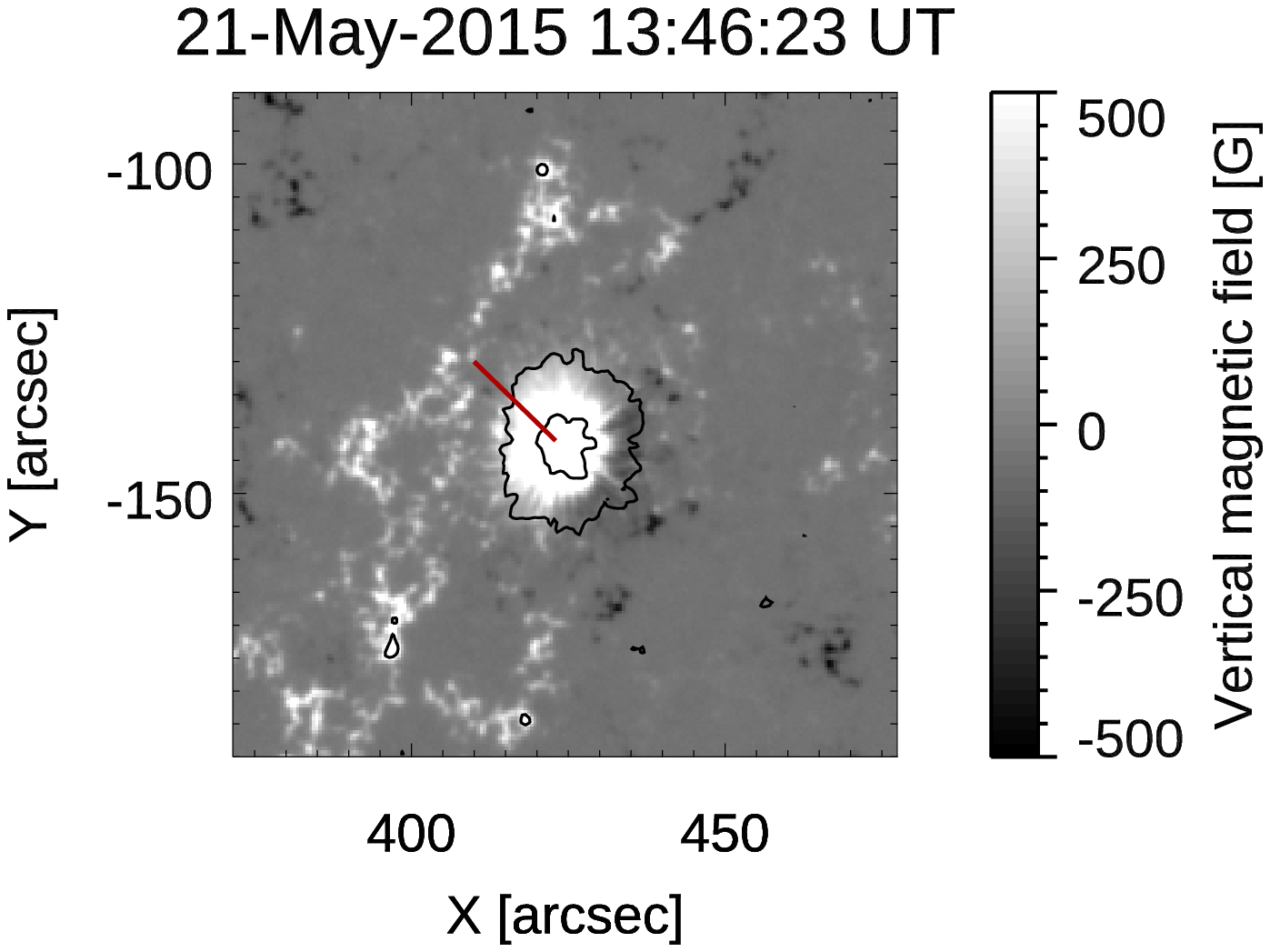}
\caption{SHARP magnetograms of AR NOAA 12348 acquired by HMI/SDO. The red circles in the top right panel indicates the location of the observed MMFs. Here and in the following Figures the inner and the outer contours correspond to the umbra-penumbra and penumbra-quiet Sun boundaries, respectively. The red segment in the bottom right panel indicates the radial cut where the magnetic field strength variation reported in the top right panel of Figure \ref{fig9} has been computed. More details can be seen in the accompanying online movie ($Magnetogram\_AR12348.wmv$; see main text). \label{fig2}} 
\end{center}
\end{figure}

During the disruption of the penumbral filaments the sunspot was also surrounded by moat flows (see the online movie accompanying Figure \ref{fig2}), as it usually occurs for decaying sunspots \citep{Sva14}. Most of the magnetic elements carried by the moat flow was bipolar, with the inner polarity of the same sign of the sunspot polarity, in accordance with \citet{Har73}. These features, recognized as moving magnetic features (MMFs), persist long enough to reach the surrounding network (top left panel of Figure~\ref{fig2}). However, in the sector corresponding to the disrupting portion of the penumbra, we observed MMFs characterized by an opposite orientation of the bipoles, i.e., with the inner polarity of opposite sign with respect to the sunspot polarity (see the red circles in the top right panel of Figure~\ref{fig2}). It seems that the outer positive components of those MMFs added flux to the positive magnetic field of the fragmented portion of the sunspot (bottom left panel of Figure~\ref{fig2}). 

\begin{figure}
\begin{center}
\includegraphics[trim=80 180 10 230, clip, scale=0.6]{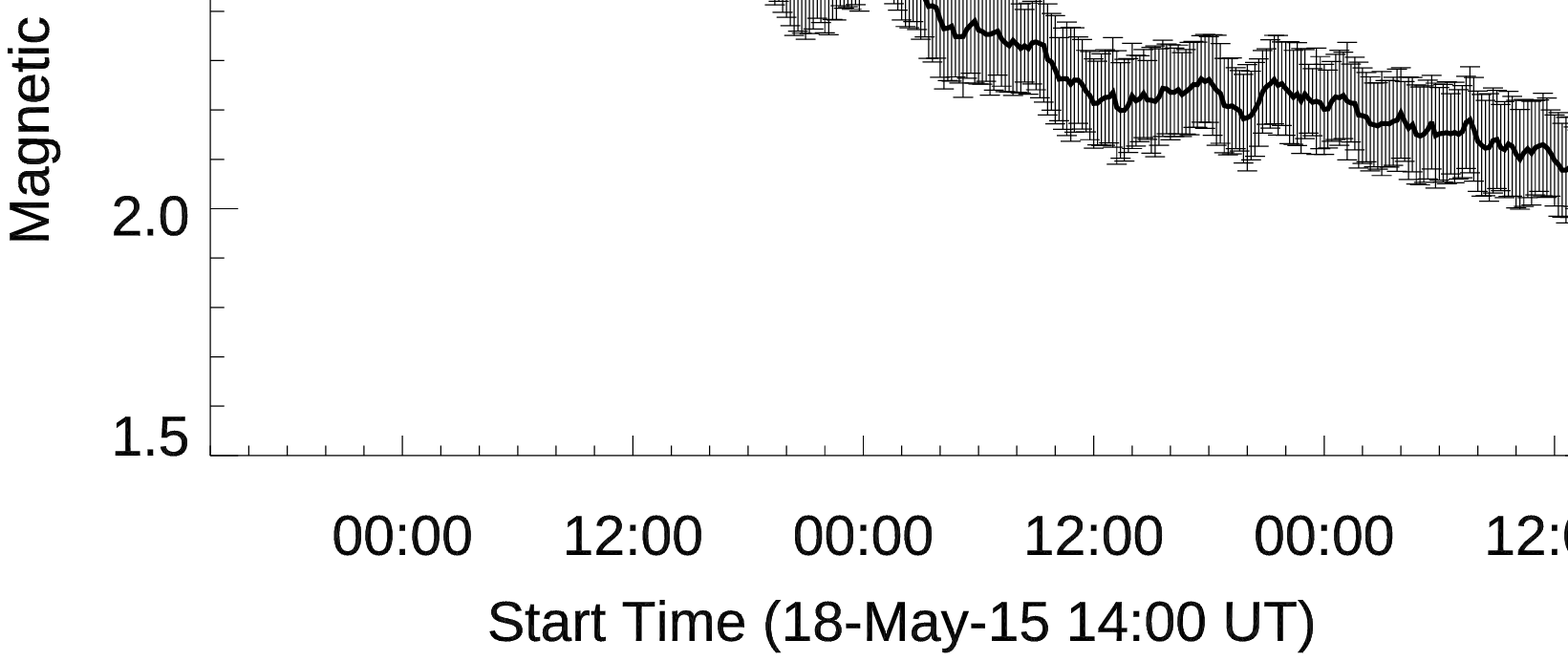}
\caption{Evolution of the magnetic flux computed using the SHARP magnetograms over the FOV indicated by the black box in the top left panel of Figure~\ref{fig1} and corresponding to the IBIS FOV. \label{fig2bis}} 
\end{center}
\end{figure}

The total magnetic flux of the whole sunspot, computed in the box of the top left panel of Figure~\ref{fig1}, exhibits a monotonic decrease during the considered time interval of HMI observations, confirming that the sunspot was in a decay phase (Figure~\ref{fig2bis}). We remark that this decrease is within the errors on the magnetic flux, estimated by propagating the experimental errors and considering the HMI sensitivity of 10~G \citep{Scho12}. 
We note that this estimation of the magnetic flux takes into account projection effects by considering the cosine of the heliocentric angle in the effective pixel size. A more sophisticate correction, like that proposed by \citet{Falconer16} for HMI/SDO data, would be required if the sunspot were observed at longitude angles larger than $\pm 45^{\circ}$.

Therefore, we can summarize several aspects that contribute to the decay of the sunspot: the disappearance of its penumbra, the reduction of its size, the presence of MMFs, and the decrease of the magnetic flux. It is noteworthy that the moat flow was not visible anymore during the restoring phase of the penumbra (bottom right panel of Figure~\ref{fig2}).

\begin{figure}
\begin{center}
\includegraphics[trim=5 80 280 430, clip, scale=0.6]{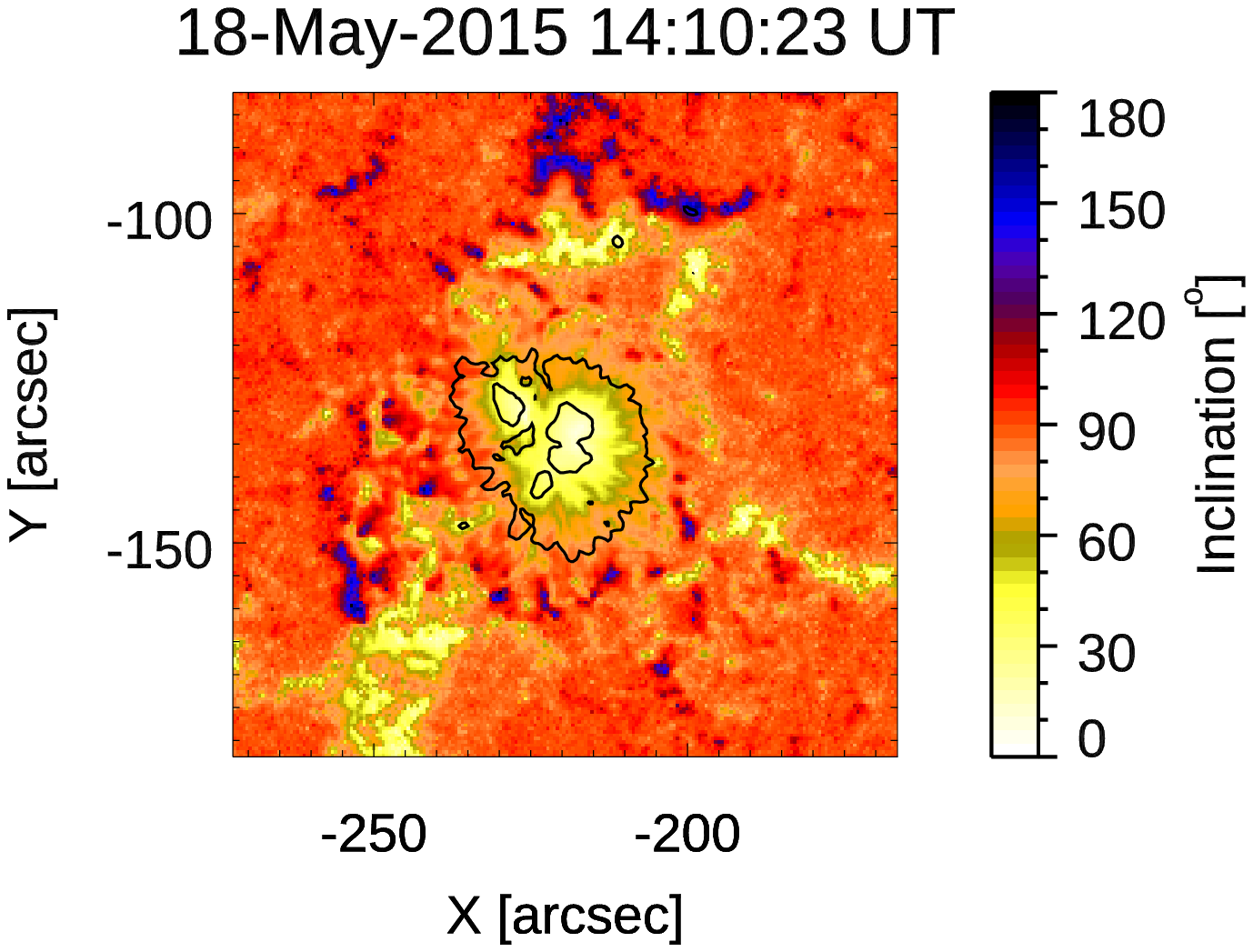}
\includegraphics[trim=35 80 100 430, clip, scale=0.6]{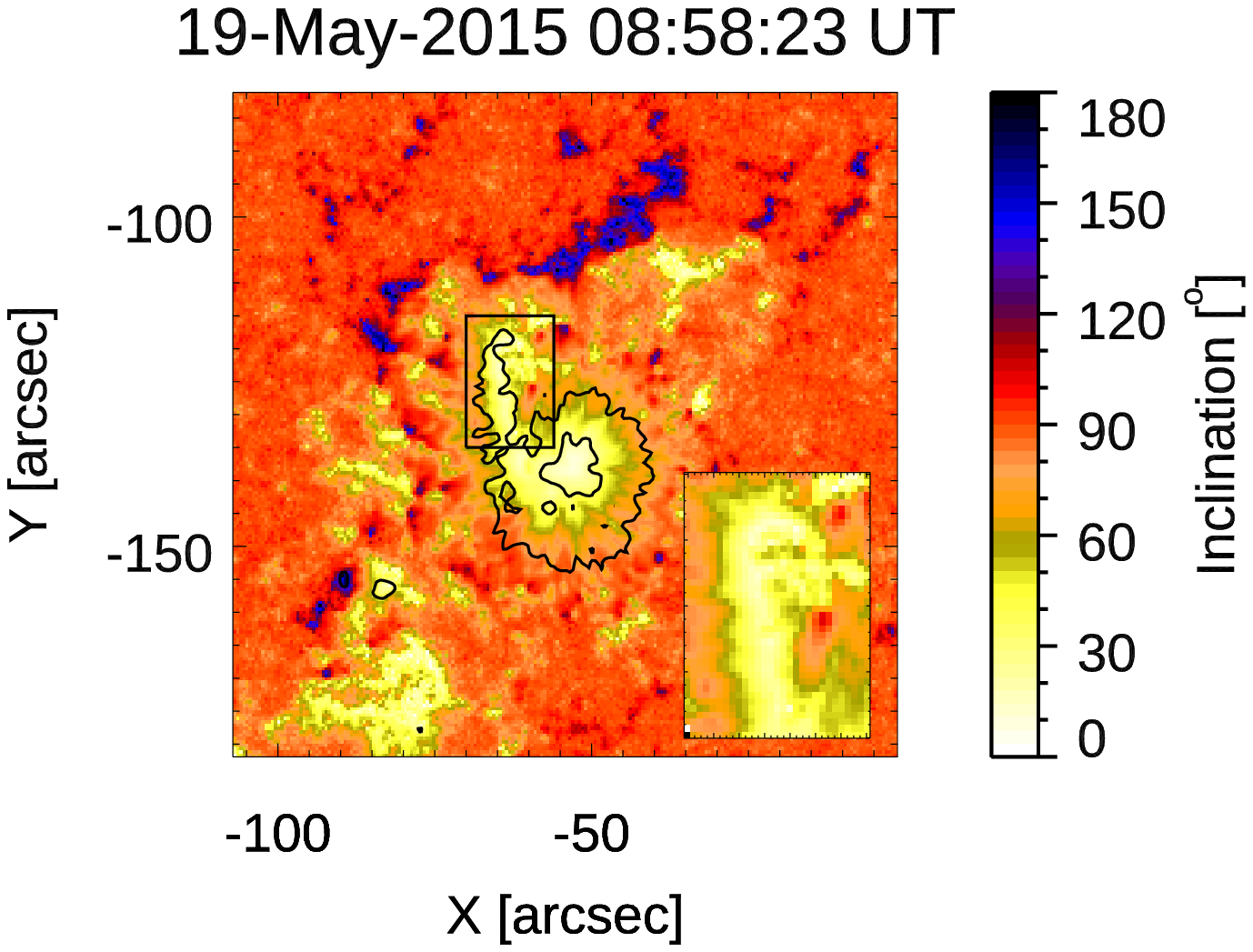}\\
\includegraphics[trim=5 80 280 430, clip, scale=0.6]{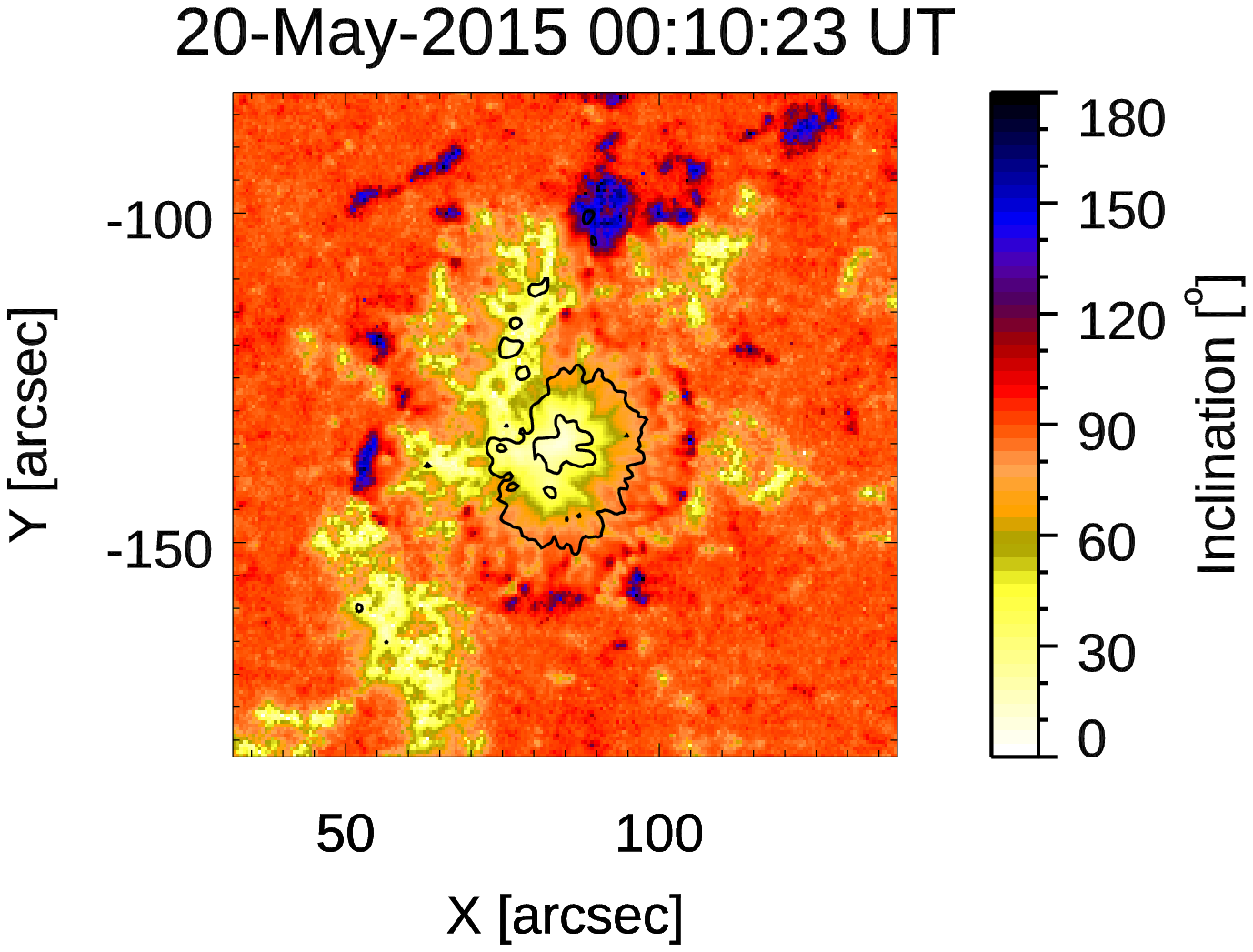}
\includegraphics[trim=35 80 100 430, clip, scale=0.6]{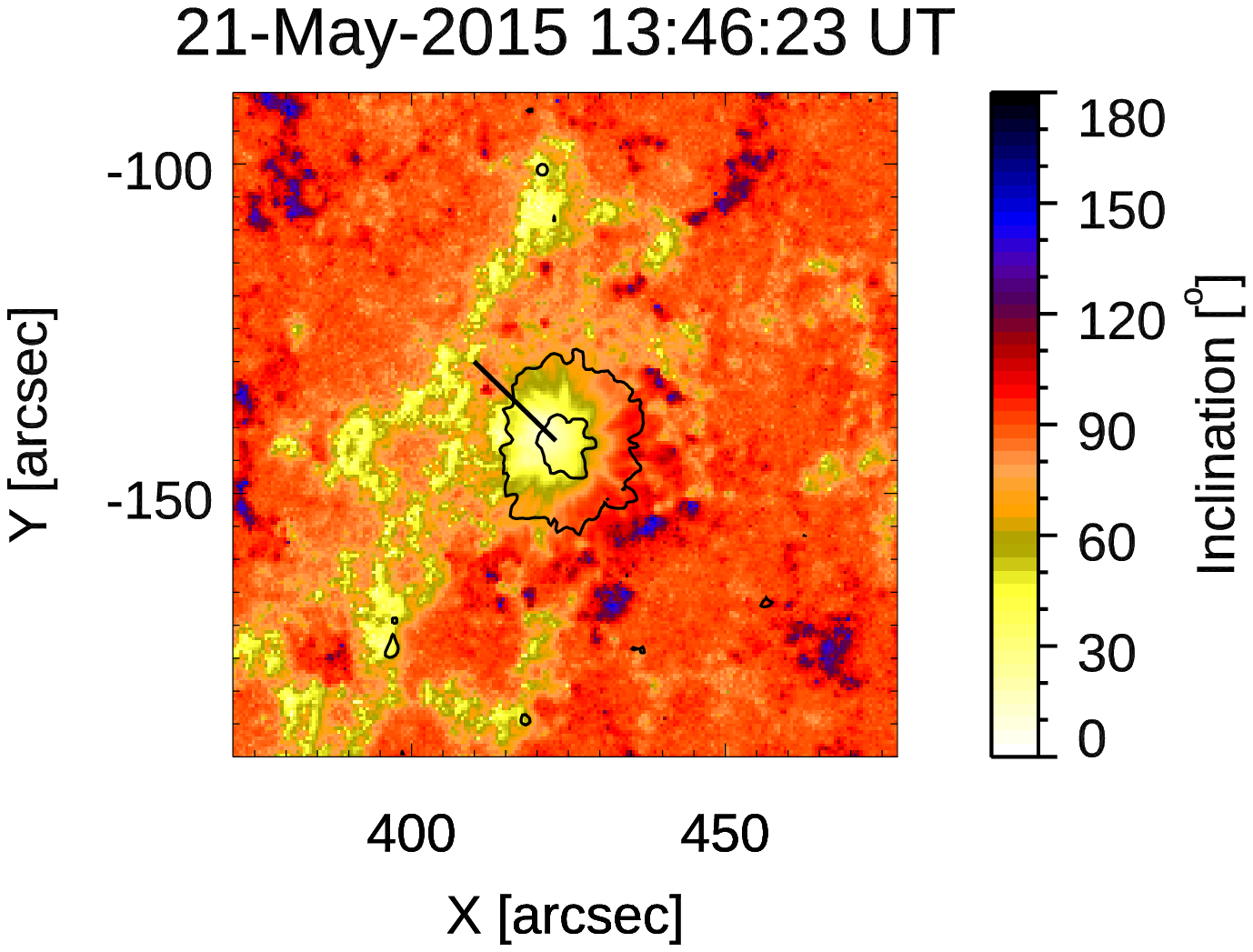}
\caption{Inclination maps of AR NOAA 12348 obtained by HMI/SDO data. The segment in the bottom right panel indicates the radial cut where the inclination variation reported in the bottom left panel of Figure~\ref{fig9} has been computed. More details can be seen in the accompanying online movie ($Inclination\_AR12348.wmv$; see main text). \label{fig3}} 
\end{center}
\end{figure}

The maps of the magnetic field inclination (Figure~\ref{fig3} and the accompanying online movie $Inclination\_AR12348.wmv$) highlight some interesting results about the settlement of the new portion of the penumbra. We clearly see the correspondence between the positive and negative polarities (Figure~\ref{fig2}) and the upward and downward directions of the magnetic field lines (Figure~\ref{fig3}), respectively. During the first time interval of observations we note a continuous northward displacement of the elongated portion of the sunspot characterized by an inclination less than 45$^{\circ}$ (top panels of Figure~\ref{fig3}). When this portion detached from the main sunspot, it contributed to the network field (bottom left panel of Figure~\ref{fig3}). In the inclination maps obtained from May 20 till the end of the observation time interval we see in the Northern part of the sunspot the formation of new filamentary structures characterized by an inclination similar to the other sectors of the penumbra. This evolution of the magnetic field inclination seems to occur progressively outwards from the edge between the umbra and the quiet Sun. The patches characterized by an inclination larger than 150$^{\circ}$, and probably corresponding to the tail of the penumbral filaments, appeared only after the restored penumbra sector was visible in the continuum filtergrams (compare the bottom right panels of Figures~\ref{fig1} and~\ref{fig3}).

\begin{figure}
\begin{center}
\includegraphics[trim=5 80 280 430, clip, scale=0.6]{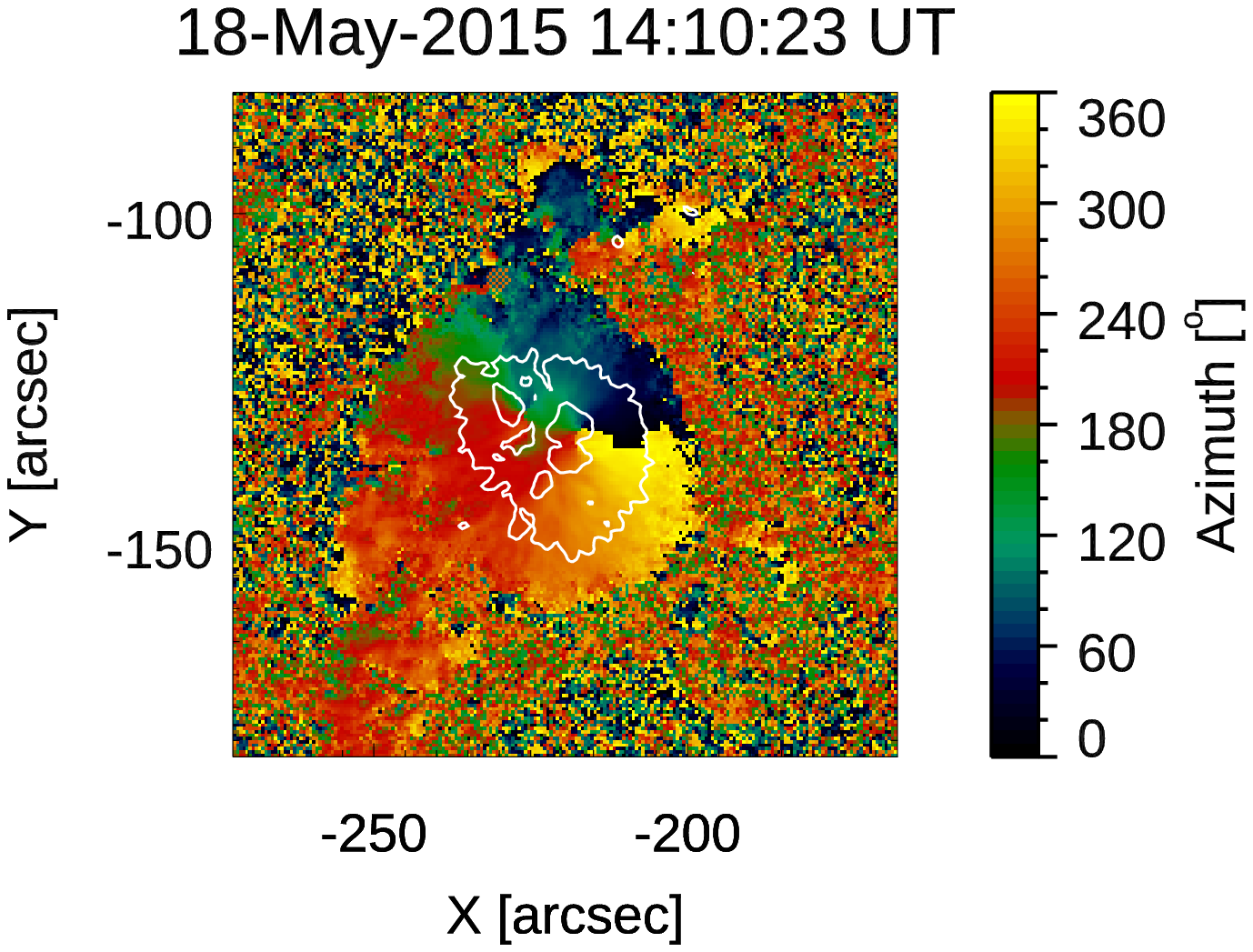}
\includegraphics[trim=45 80 100 430, clip, scale=0.6]{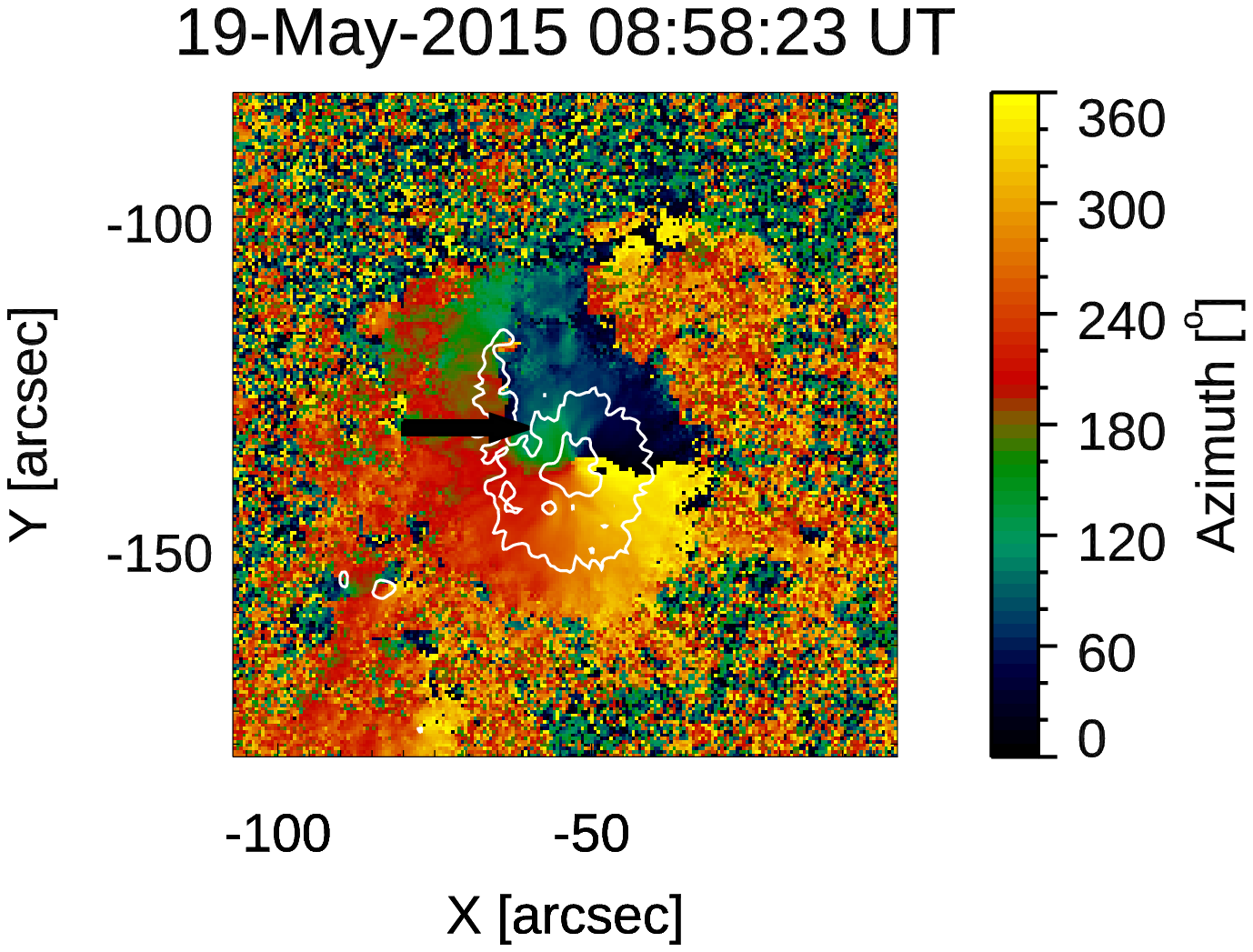}\\
\includegraphics[trim=5 80 280 430, clip, scale=0.6]{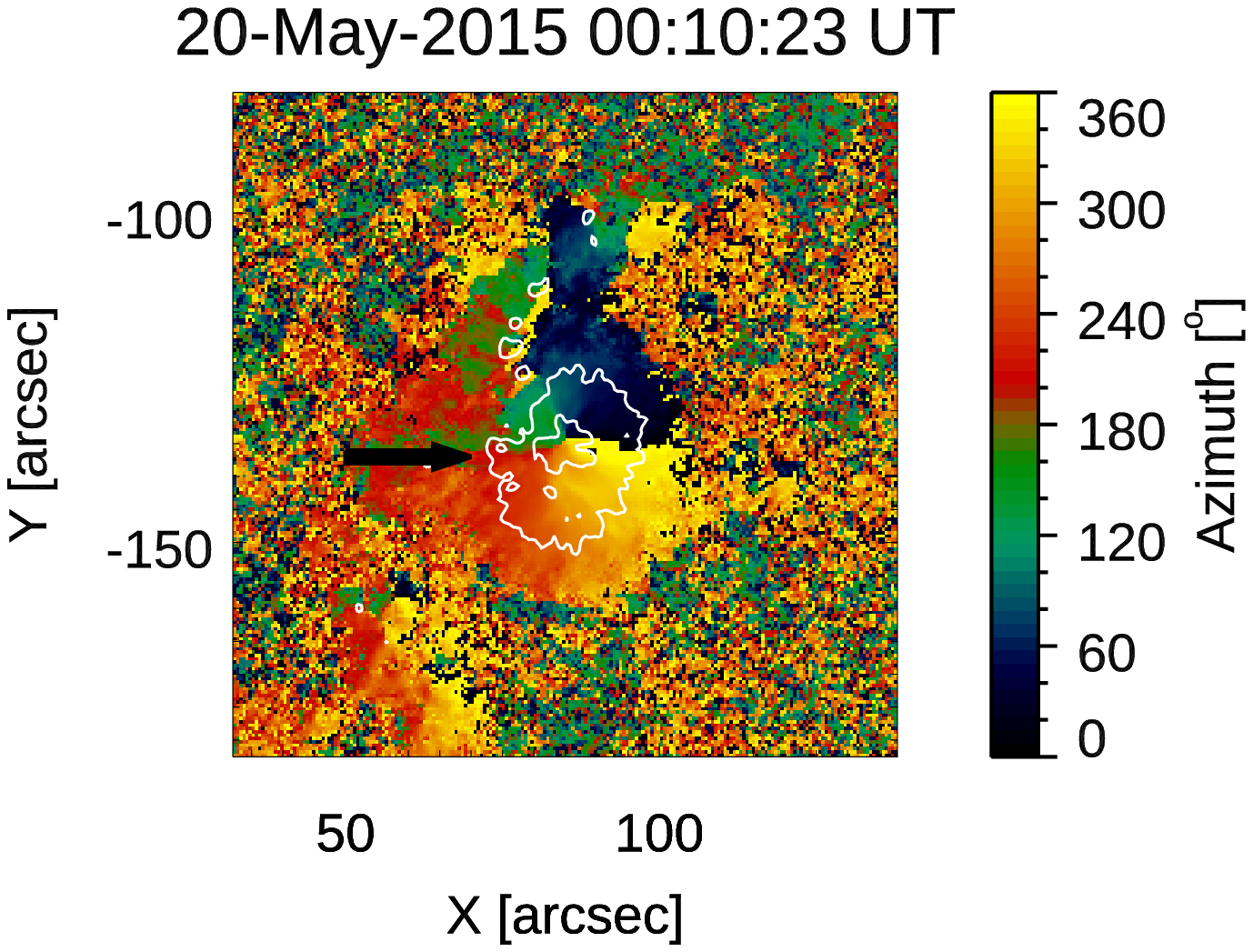}
\includegraphics[trim=45 80 100 430, clip, scale=0.6]{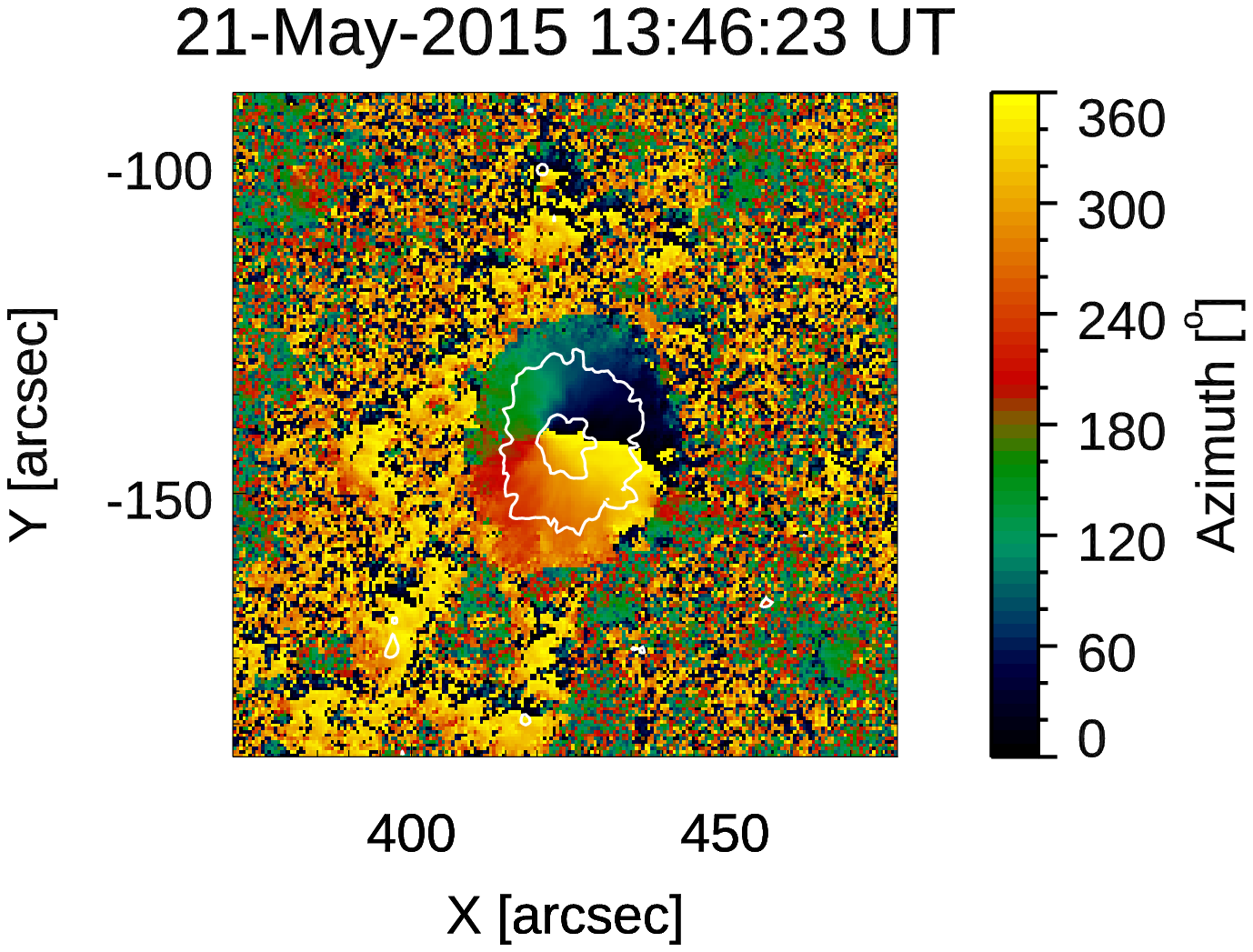}
\caption{Azimuth maps of AR NOAA 12348 obtained by HMI/SDO data. More details can be seen in the accompanying online movie ($Azimuth\_AR12348.wmv$; see main text). \label{fig4}} 
\end{center}
\end{figure}

We also analyze the azimuth angle maps relevant to the sunspot, to highlight its evolution during penumbra disappearance and restoring. Indeed, a fully-formed sunspot with a regular configuration shows a smooth transition of the azimuth angle values between 0$^{\circ}$ and 360$^{\circ}$ around it \citep{Sol03}. In our observations, a discontinuity in the radial distribution of the penumbra is visible at the beginning of the observations (top right panel of Figure \ref{fig4}), when a green sector corresponding to azimuth angle values between 90$^{\circ}$ and 150$^{\circ}$ is being interrupted by the intrusion of a Westward field, corresponding to the LB visible in the top left panel of Figure~\ref{fig1} (see also the continuum contours). The restoration of the penumbral sector seems to proceed from the inner part of the sunspot towards the outer part also in the azimuth maps: from May 20 (bottom left panel of Figure~\ref{fig4}) to May 21 (bottom right panel of Figure~\ref{fig4}) we note that the portion of the magnetic field with an azimuth angle between 90$^{\circ}$ and 150$^{\circ}$ (green in Figure~\ref{fig4}) increased its size in time, reaching the edge of an ideal circle surrounding the sunspot when the penumbra is completely restored.

\subsection{High spatial resolution observations of the penumbra carried out by IBIS}

\begin{figure}
\begin{center}
\includegraphics[trim=180 330 160 260, clip, scale=0.65]{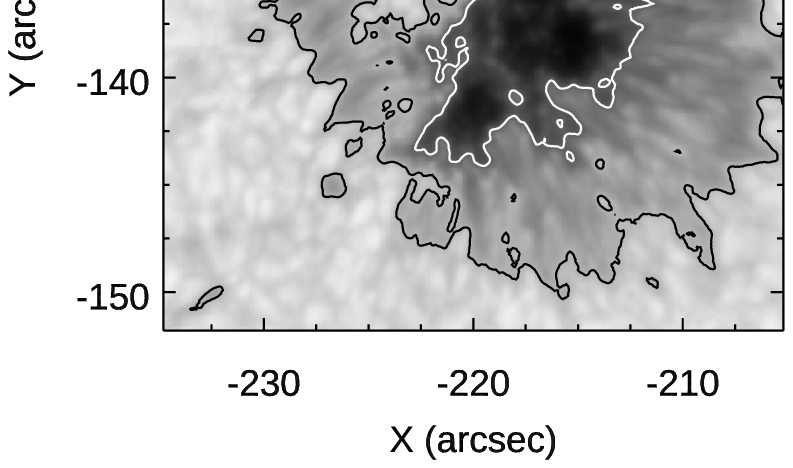}
\includegraphics[trim=205 330 50 260, clip, scale=0.65]{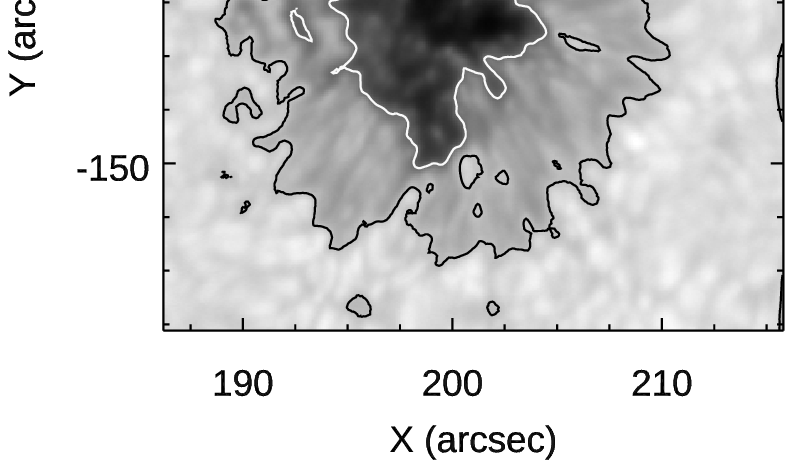}\\
\includegraphics[trim=180 330 160 260, clip, scale=0.65]{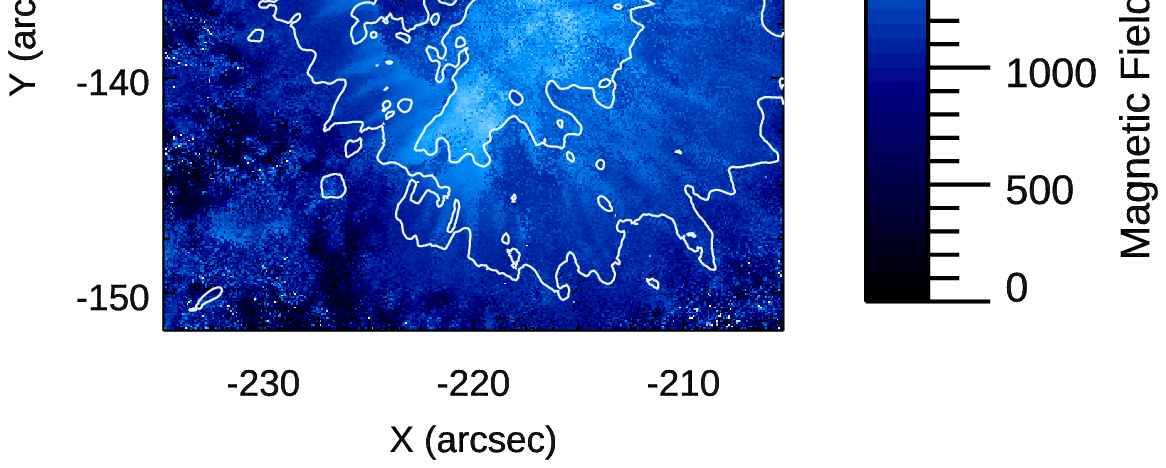}
\includegraphics[trim=205 330 50 260, clip, scale=0.65]{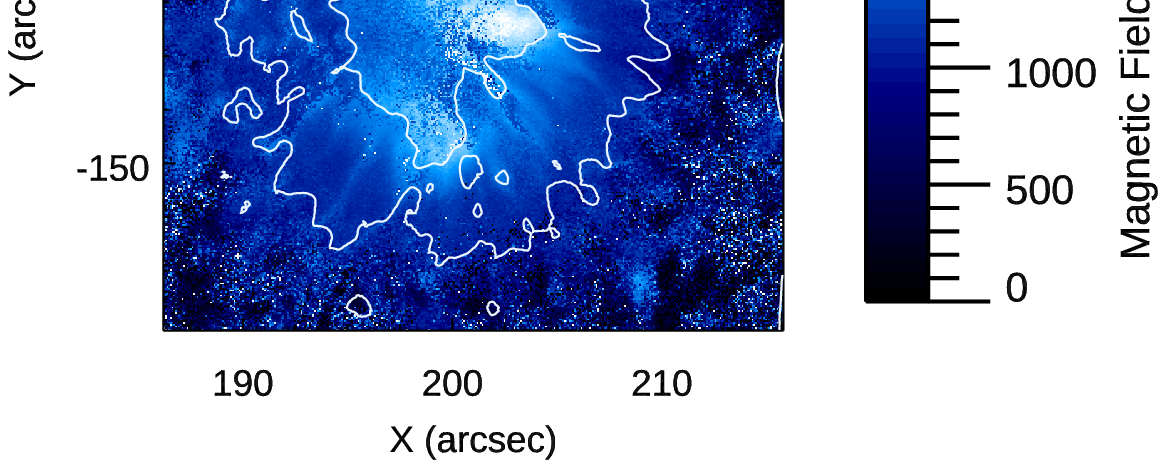}\\
\includegraphics[trim=180 300 160 260, clip, scale=0.65]{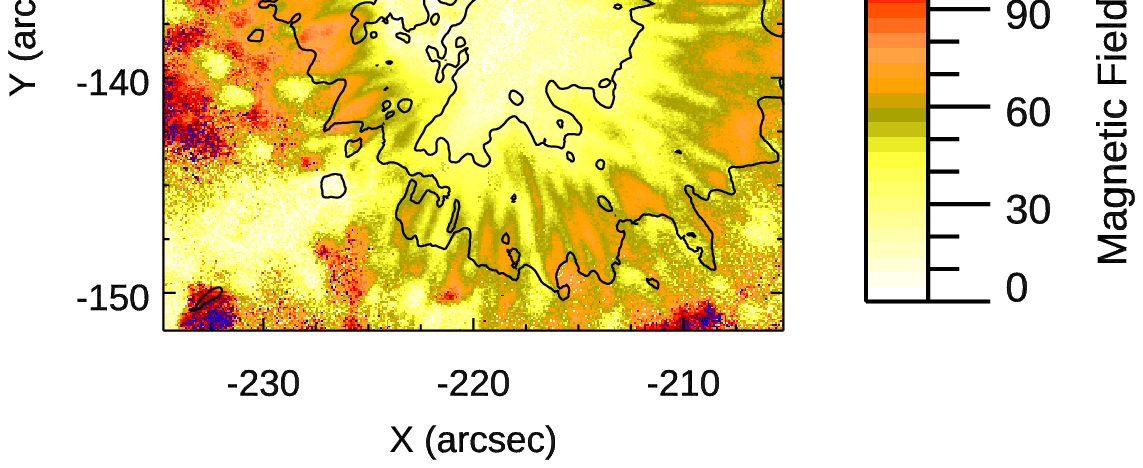}
\includegraphics[trim=205 300 50 260, clip, scale=0.65]{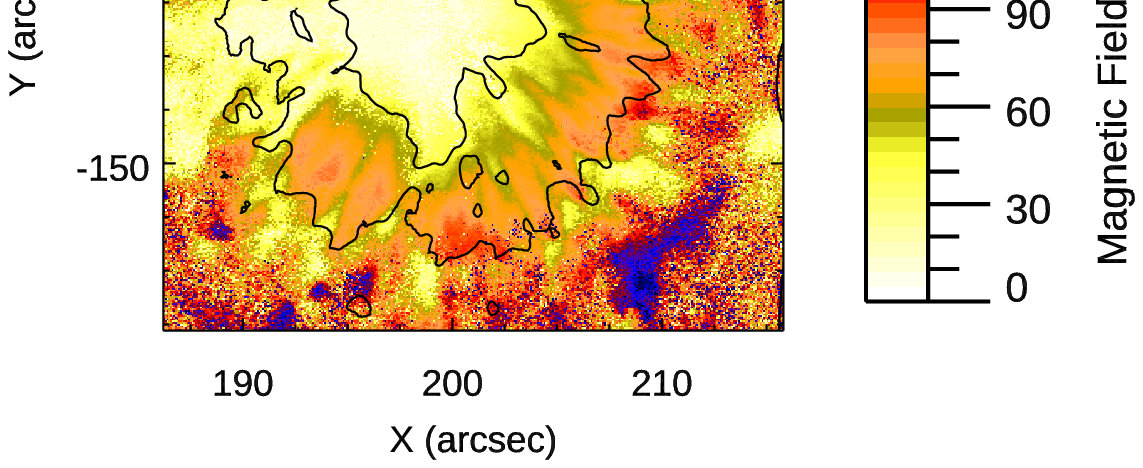}
\caption{Maps of the intensity (top row), magnetic field strength (middle row), and inclination angle (bottom row) obtained from the SIR inversion of the Stokes profiles of the \ion{Fe}{1} 630.25~nm line (IBIS dataset). The inner and the outer contours indicate the umbra-penumbra and the penumbra-quiet Sun boundaries, respectively. The surface normal is the reference for the inclination angle. The boxes drawn in the left panels indicate the region enlarged in Figure~\ref{fig5bis}. \label{fig5}} 
\end{center}
\end{figure}

\begin{figure}
\begin{center}
\includegraphics[trim=170 300 50 260, clip, scale=0.7]{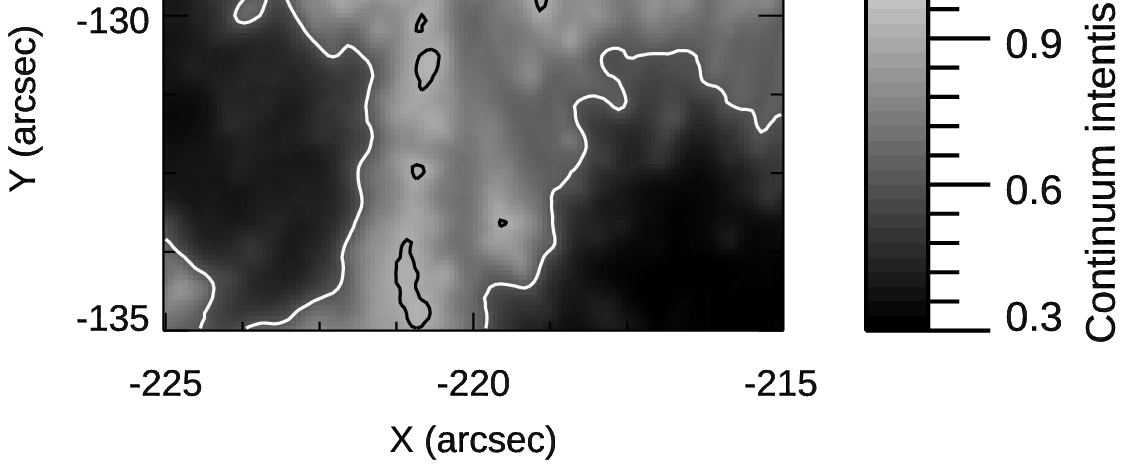}\\
\includegraphics[trim=170 300 50 372, clip, scale=0.7]{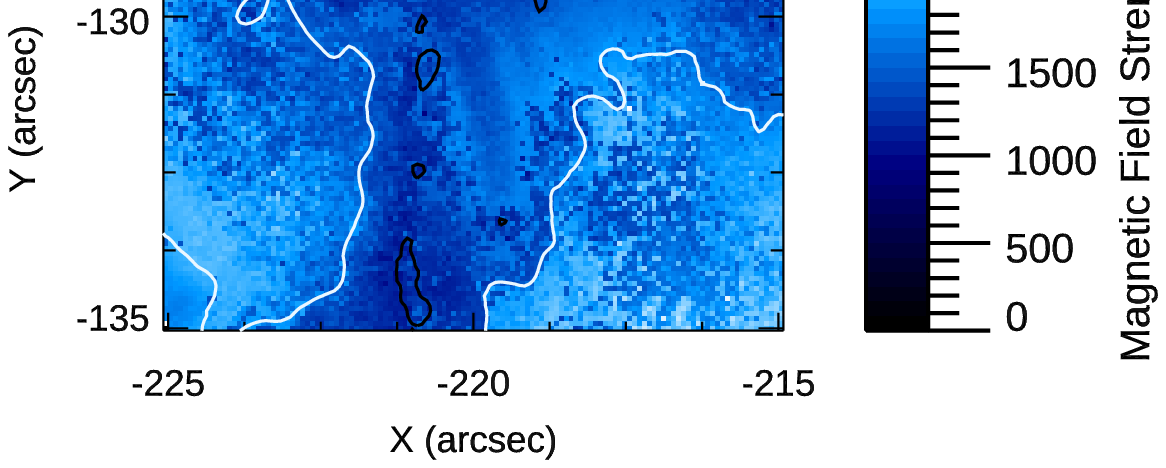}\\
\includegraphics[trim=170 300 50 372, clip, scale=0.7]{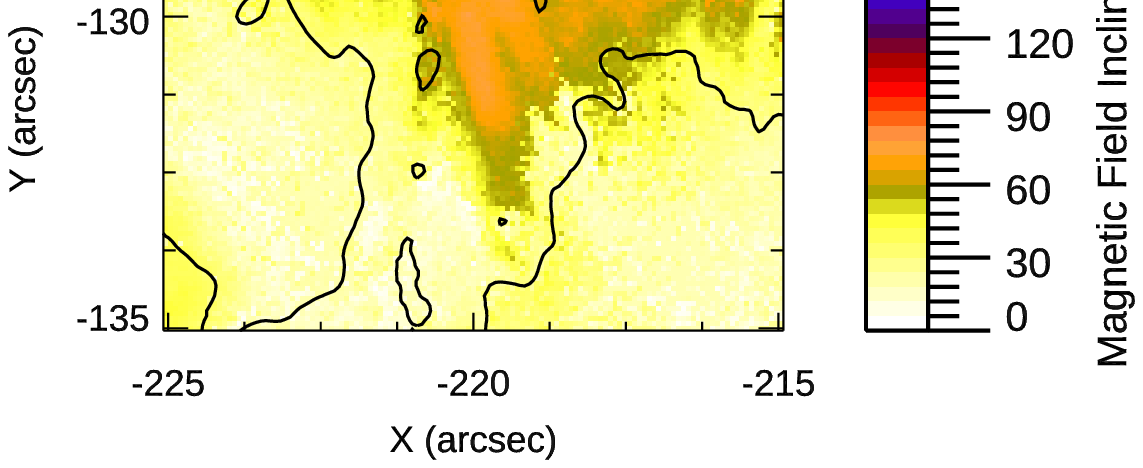}\\
\caption{Zoomed images inside the box drawn in the left panels of Figure~\ref{fig5}. From top to bottom: continuum intensity, magnetic field strength, and inclination angle maps obtained by the inversion of IBIS spectro-polarimetric measurements. \label{fig5bis}} 
\end{center}
\end{figure}

The high-resolution spectro-polarimetric images taken by IBIS allow us to study in detail the properties of the magnetic field during the decay and restoring phases of the penumbral sector. Nevertheless, only few scans were acquired during best seeing conditions on May 18 and 20, so that we are not able to reconstruct the evolution of the target during all crucial phases using IBIS data, we can infer the magnetic field configuration in detail during the penumbra decay and during the penumbra restoring process. 

On May 18, the IBIS continuum images (see top panel of Figure~\ref{fig5}) showed more details of the LB. It was about 5\arcsec{} wide and no dark lane was visible at the IBIS resolution. In the Northern portion of the LB, i.e., the region inside the box, we can identify some granules surrounded by two penumbral filaments (see the zoomed continuum map in Figure~\ref{fig5bis}). This region could be due to convective motions taking over the magneto-convection among the penumbral filaments. In the map of the magnetic field strength (see the arrow in the middle left panel of Figure \ref{fig5}), we can see that the LB is characterized by a field strength of about 1000~G and surrounded by a stronger field of about 1500~G. The inclination of the magnetic field in the LB was less than 30$^{\circ}$ (bottom left panel of Figure~\ref{fig5}), indicating the presence of a magnetic field on average more vertical than in the penumbra, where the uncombed structure \citep{Sol93} was formed by spines (more vertical and stronger fields) and intra-spines (more horizontal and weaker fields) interlaced with each other \citep{Tho04}. The only region of the LB characterized by horizontal fields corresponds to its Northern portion, as shown in the inclination map of Figure~\ref{fig5bis} displaying the region inside the box of Figure~\ref{fig5}.

On May 20, the best IBIS scan was obtained at 13:38~UT, when the penumbra was going to be completely formed (see top right panel of Figure~\ref{fig5}). At that time, the umbra of the sunspot appeared neither symmetric nor homogeneous in intensity. In comparison with the data taken on May 18, we note higher strength of the magnetic field of the umbra, in its Western side (see the middle right panel of Figure~\ref{fig5}).

The continuum image also shows restored sectors of the penumbra in the same location where the magnetic field was diffused during the sunspot fragmentation. That portion of the restoring penumbra was characterized by a magnetic field strength comparable with the other sectors of the penumbra. 

Penumbral filaments characterized by an inclination of the magnetic field of about 80$^{\circ}$ along their main axis (see the arrows in the bottom panels of Figure \ref{fig5}) replaced the region of the LB where on May 18 some granules seemed surrounded by two penumbral filaments. In the map displaying the inclination of the magnetic field on May 20 we also note that the region of the photosphere where the penumbra was not yet visible in the continuum map, the magnetic field had an inclination between 20$^{\circ}$ and 40$^{\circ}$ (see the black arrow in the North-Eastern part of the FOV in the bottom right panel of Figure \ref{fig5}). Instead, a vertical field (less than 20$^{\circ}$) was observed in the region extending along the Eastern direction and connecting the umbra with the sectors of the penumbra still involved in the restoring process.

\subsection{Evershed flow evolution}

\begin{figure}
\begin{center}
\includegraphics[trim=5 200 40 30, clip, scale=0.8]{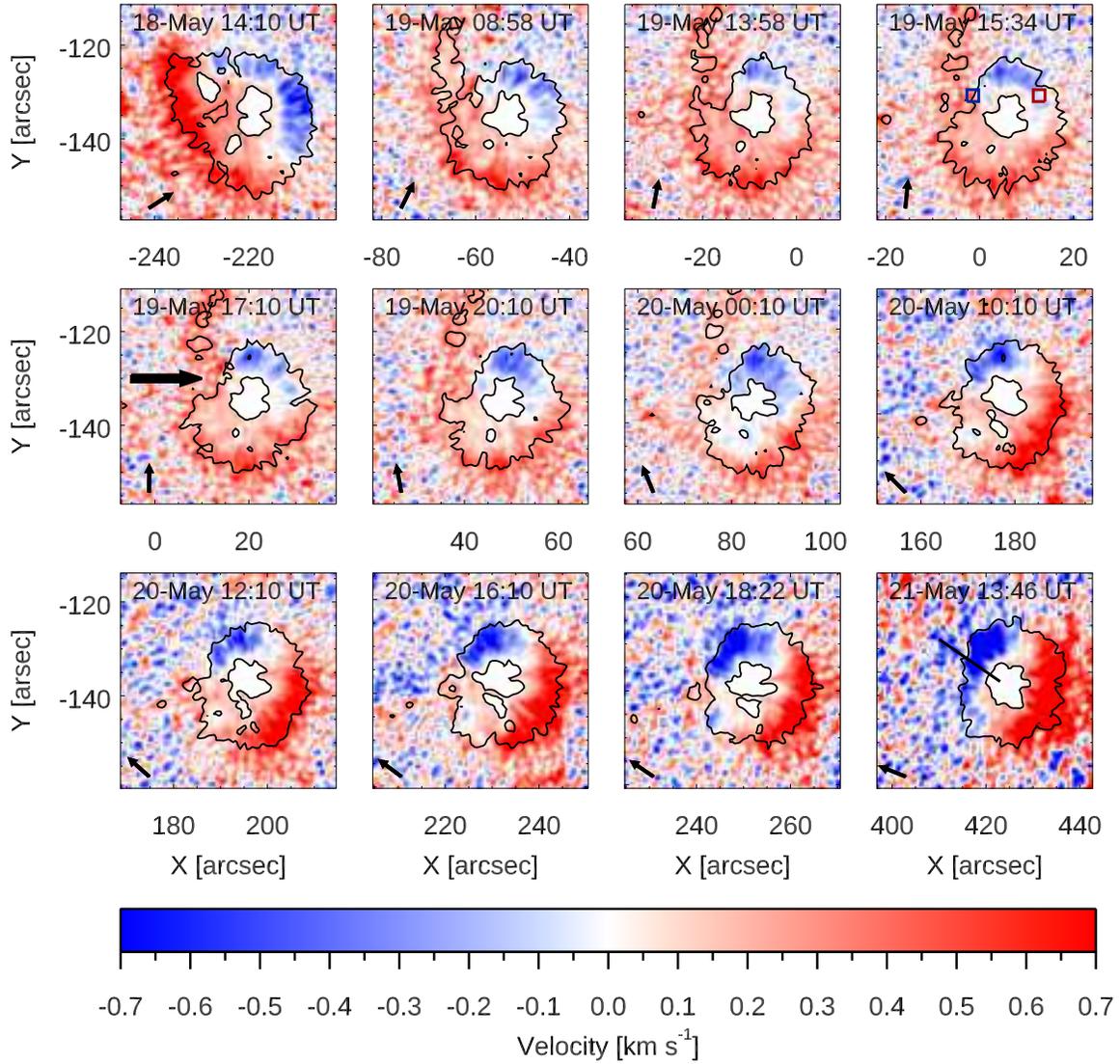}
\caption{\footnotesize Doppler maps of AR NOAA 12348 obtained by HMI/SDO dataset and showing the plasma velocity along the LOS in the same FOV of IBIS. More details can be seen in the accompanying online movie ($Doppler\_AR12348.wmv$; see text). Similarly to Figure~\ref{fig1}, the blue box in the map taken on May 20 at 16:10~UT frames the region of interest where the evolution of the LOS velocity has been measured, as plotted in the bottom right panel of Figure~\ref{fig10}. For comparison, we considered also a red box in the stable penumbra. The segment in the bottom right panel indicates the radial cut where the velocity variation reported in the bottom left panel of Figure~\ref{fig9} has been computed. The arrows in the bottom left corner of each panel point to the disk center. \label{fig7}} 
\end{center}
\end{figure}

The evolution of the Evershed flow in the portion of the penumbra involved by the restoring process confirms that we can consider the transition from counter-Evershed flow into the classical Evershed flow as a key signature of the formation of penumbral filaments \citep[see][]{Mur16}. 

From the Dopplergrams of the SHARP data, we can follow the evolution of the plasma flow during the disruption and reformation of part of the penumbra. At the beginning of the HMI dataset, when the sunspot was in the Eastern solar hemisphere, we distinguish clearly the redshift and blueshift of the plasma {in the penumbral filaments pointing to the solar limb and to the disk center,  respectively (top left panel of Figure~\ref{fig7}). On May 19, while the disruption of the penumbra was still going on, a coherent Evershed flow disappeared in correspondence of the fragmented portions of the sunspot. On May 19 from 15:00 UT the sector of the sunspot involved by the penumbra reformation (azimuth angles from $120^{\circ}$ to $210^{\circ}$) was characterized by the presence of the counter-Evershed flow, which is typical of the formation phases of penumbral filaments \citep[e.g.,][]{Mur16}. Indeed, as can be seen from the middle row panels of Figure~\ref{fig7}, in particular we find redshift toward the disk center at azimuth angles $135^{\circ}-180^{\circ}$, in contrast to the blueshift one would expect %due to the change of location of the sunspot 
(see, e.g., the arrow in the left panel of the middle row of Figure~\ref{fig7}).
In the bottom row of Figure~\ref{fig7}, we still see that the counter-Evershed flow is observed at 12:10~UT at azimuth angles $180^{\circ}-210^{\circ}$, being sequentially replaced along the counterclockwise direction by upflow toward the disk center (classical Evershed flow). This means that also the change from counter-Evershed flow to classical Evershed flow occurred from azimuth angles $120^{\circ} - 210^{\circ}$, along the same counterclockwise direction as the penumbra reformation observed in the continuum intensity maps. Finally, on May 21 when the penumbra was fully restored (bottom right panel of Figure~\ref{fig7}), the flows around the sunspot are arranged accordingly to the classical Evershed flow}.

The online movie of the Doppler maps ($Doppler\_AR12348.wmv$) shows better the starting phase of the classical Evershed flow during the settlement of the penumbral filaments on May 20. The Evershed flow was completely re-established on May 21 (bottom right panel of Figure~\ref{fig7}). Due to the fact that the restoring process of sunspot penumbra occurred during the passage of the sunspot across the central meridian, when the Evershed flow was less evident, by the Dopplergrams we are not able to estimate with enough precision the time interval necessary for the transition from the counter to classical Evershed flow.

\subsection{Parameters evolution during penumbra reformation}

{ We measured the evolution in time of continuum intensity, magnetic field strength, inclination, and LOS velocity in the region characterized by the reformation of the penumbra, estimating the average values of these quantities within a blue box, as shown in the bottom right panel of Figure~\ref{fig1}. This blue box of $5 \times 5$ pixels has been chosen in the Eastern side of the sunspot, where the penumbra disappeared and restored. In order to make a comparison of those parameters with respect to a region where the penumbra remained stable, we also selected another region of $5 \times 5$ pixels marked by a red box in the bottom right panel of Figure~\ref{fig1}. The same boxes have been indicated in Figure~\ref{fig7}.

The decrease of the continuum intensity (see the blue line in the top left panel of Figure~\ref{fig10}) corresponds to the formation of the penumbral filaments, which started on May 20 at around 00:00~UT. In about 12 hours, the continuum intensity passed from 0.9 $I_{c}$ to 0.7 $I_{c}$, i.e., to the same level of the region where the penumbra remained stable.

The magnetic field strength decreases till May 20 at around 14:00~UT, this reflects the decay phase of the AR, although an increase of about 100~G is visible after the beginning of the penumbra reformation (top right panel of Figure~\ref{fig10}).

The selected area inside the region where the penumbra reformed shows only a slight variation of the magnetic field inclination, i.e., between $25^{\circ}$ and $50^{\circ}$ (blue line in the bottom left panel of Figure~\ref{fig10}).

\begin{figure}
\begin{center}
\includegraphics[trim=60 280 0 230, clip, scale=0.425]{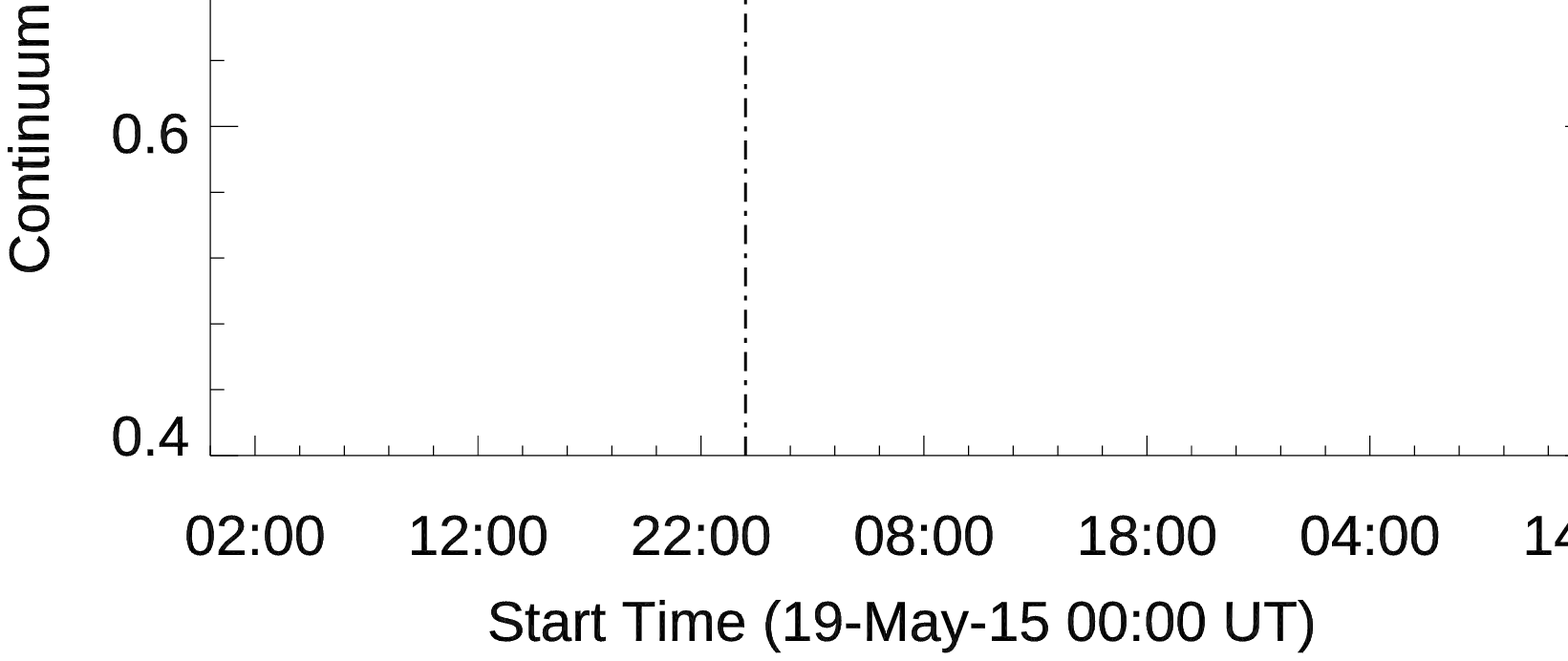}
\includegraphics[trim=60 280 0 230, clip, scale=0.425]{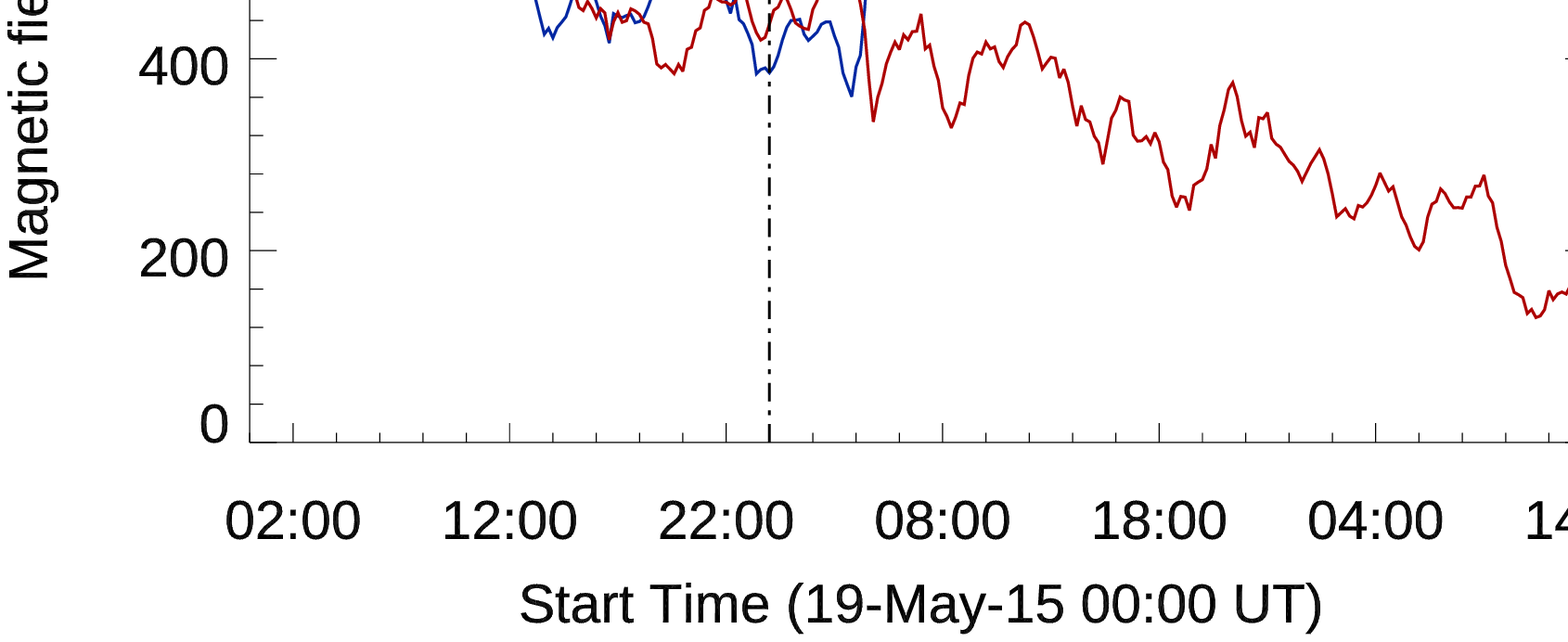}\\
\includegraphics[trim=60 180 0 230, clip, scale=0.425]{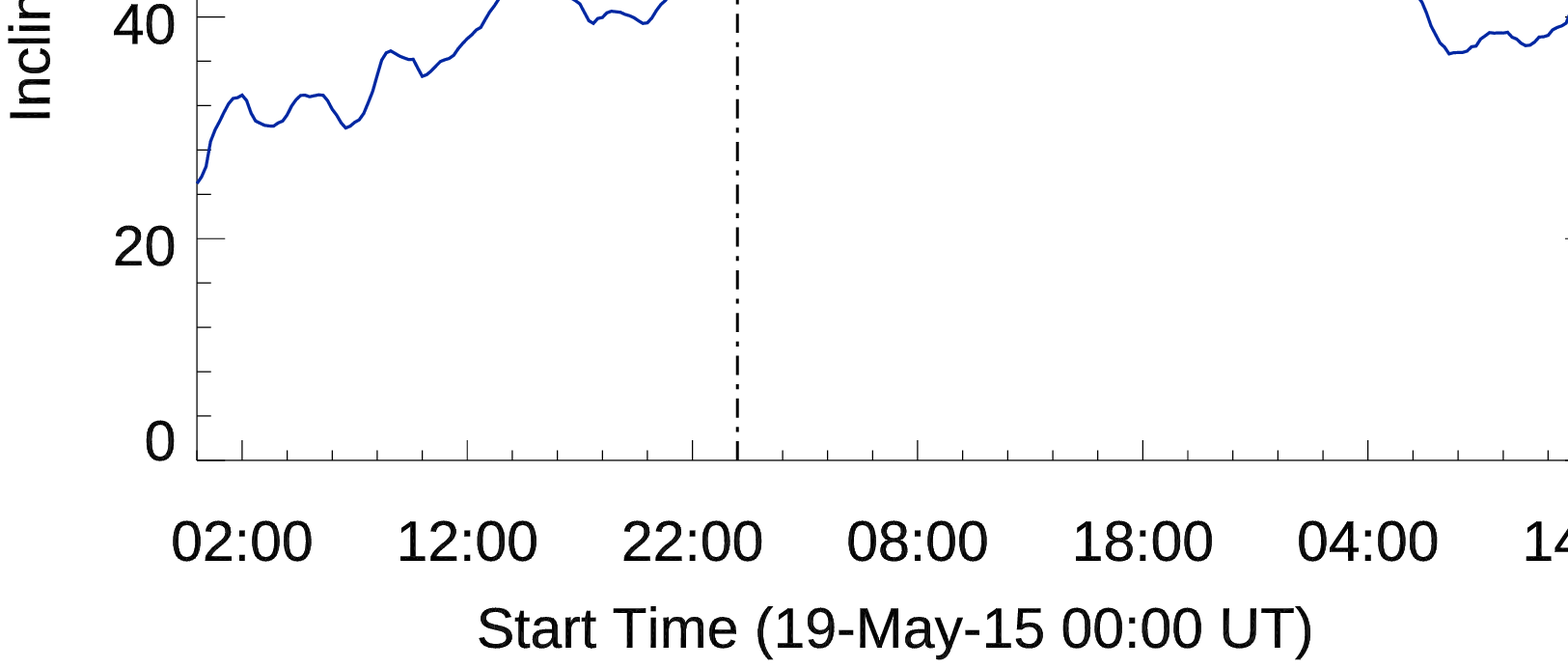}
\includegraphics[trim=60 180 0 230, clip, scale=0.425]{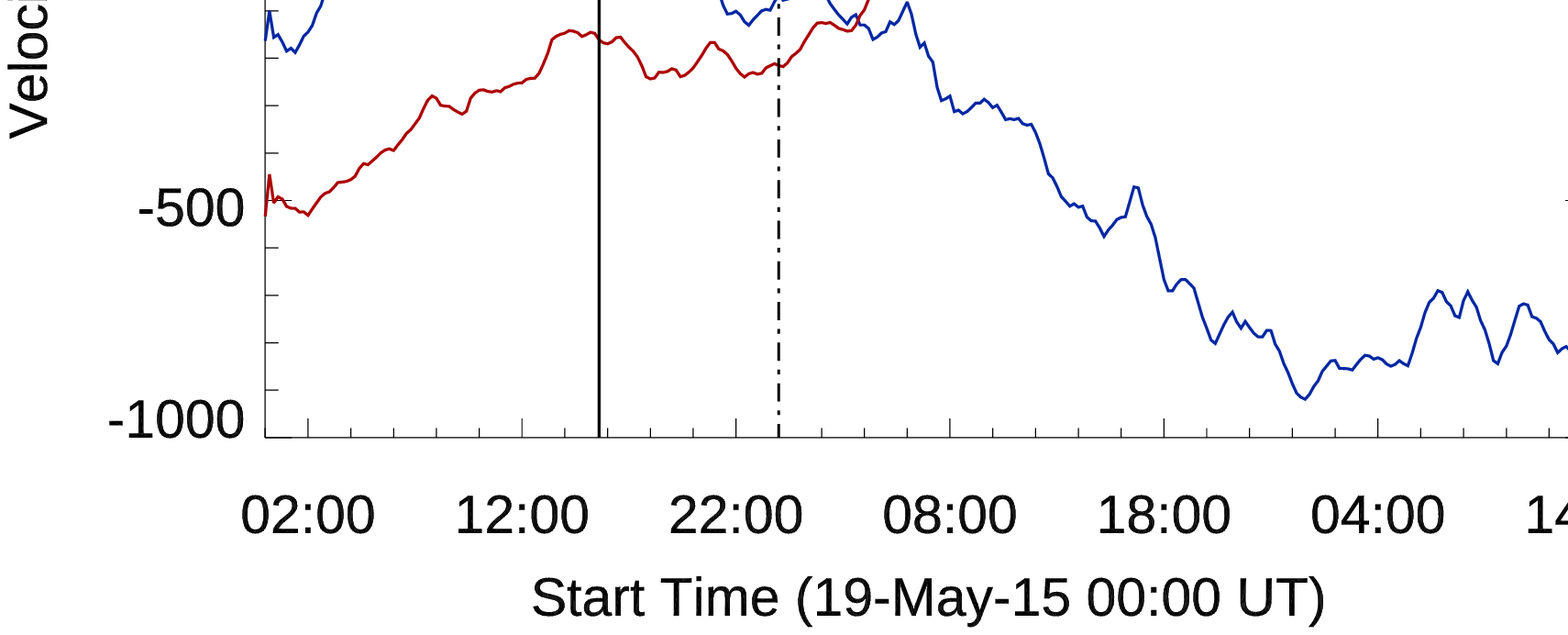}
\caption{ Evolution of the continuum intensity (top left panel), magnetic field strength (top right panel), magnetic field inclination (bottom left panel) and LOS velocity (bottom right panel), as measured in the red and blue boxes indicated in the bottom right panel of Figure~\ref{fig1}. The line colors correspond to the box colors. The vertical dash-dotted line indicates approximately the time when the penumbra started to reform. The vertical and horizontal solid lines in the bottom right panel indicate the passage of the sunspot through the central meridian and the velocity value equal to zero, respectively. \label{fig10}} 
\end{center}
\end{figure}

The counter-Evershed flow in the region of the penumbra reformation is also visible in the bottom right panel of Figure~\ref{fig10}, where the blue and red lines indicate the LOS velocity evolution measured in the areas marked by the blue and red boxes in the left panel of the middle row of Figure~\ref{fig7}. We highlight that, when the sunspot crossed the central meridian, at around 16:00~UT on May 19, as indicated by the vertical solid line, the two regions of interest were characterized by velocities with opposite directions of the LOS plasma motion. (See also the top right panel of Figure~\ref{fig7} for comparison in the maps). These opposite signs in velocity correspond to different Evershed flows, in spite of the fact that the two regions have almost the same position with respect to the disk center. In particular, the downflow corresponds to inward plasma motions along the radial filaments of the reforming penumbra (counter-Evershed flow). 

In order to describe the evolution of these parameters as a function of their distance from the sunspot center, we also plot their evolution along the radial cut marked by the segment in the bottom right panels of Figures~\ref{fig1}, \ref{fig2}, \ref{fig3} and~\ref{fig7}.} We selected four frames during the restoring process in order to better highlight the main variations of those parameters. In the plot of the continuum intensity, i.e., in the top left panel of Figure~\ref{fig9}, we see the progressive reformation of the penumbral sector by the displacement of the quiet Sun intensity level outwards from the sunspot center. In particular, on May 20 at 11:58~UT we are able to distinguish already part of the region indicated by the segment with the continuum intensity at around 0.7~$I_{c}$, corresponding to a first photospheric signature of the penumbra, whose extension reaches about 5\arcsec{} on May 21 at 13:46~UT (see the black line in Figure \ref{fig9}). Therefore, from the evolution of the continuum intensity it seems that the appearance of the penumbral sector in photosphere proceeds from the inner to outer part of the sunspot.

\begin{figure}
\begin{center}
\includegraphics[trim=35 10 10 30, clip, scale=0.3]{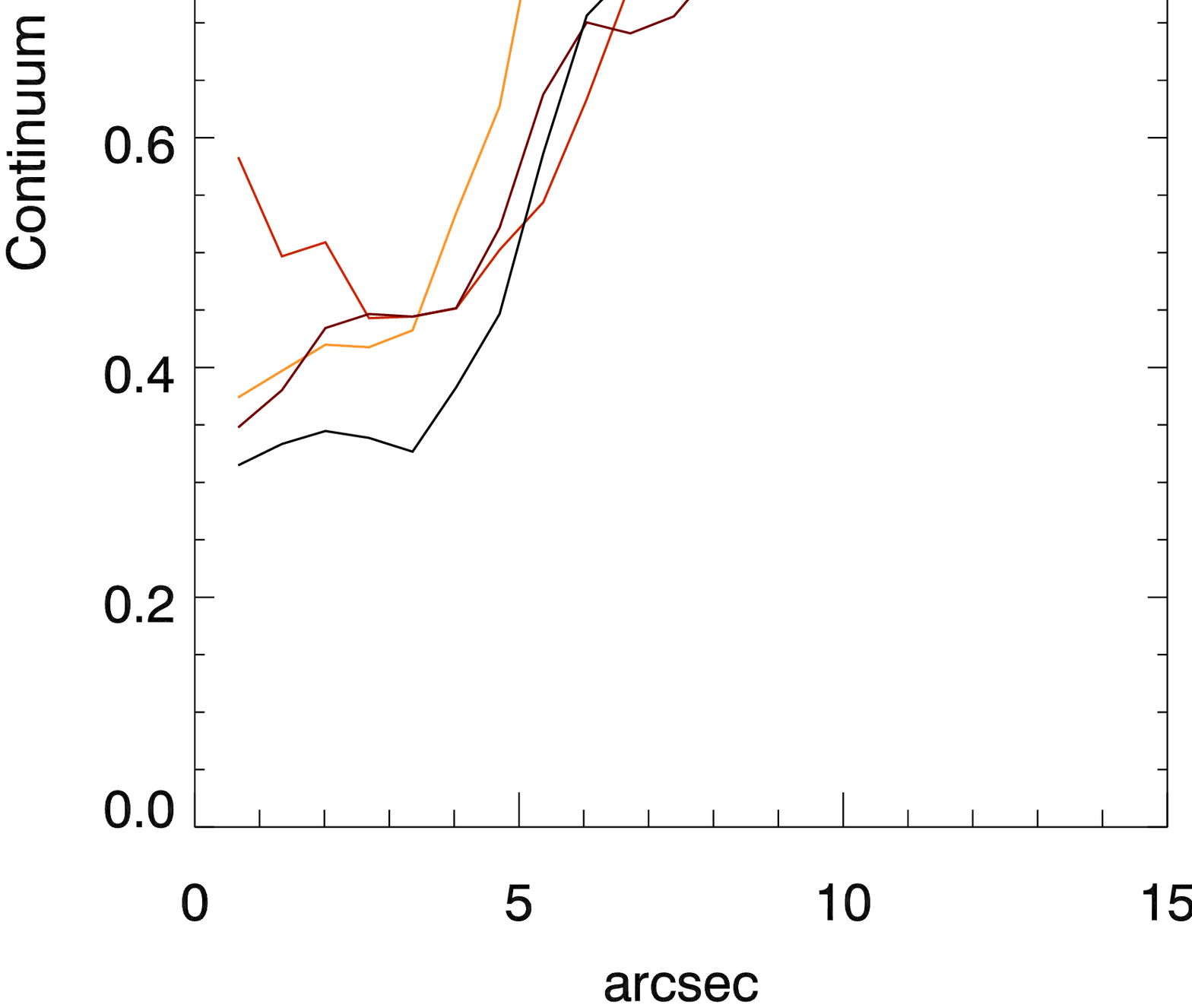}
\includegraphics[trim=35 10 10 30, clip, scale=0.3]{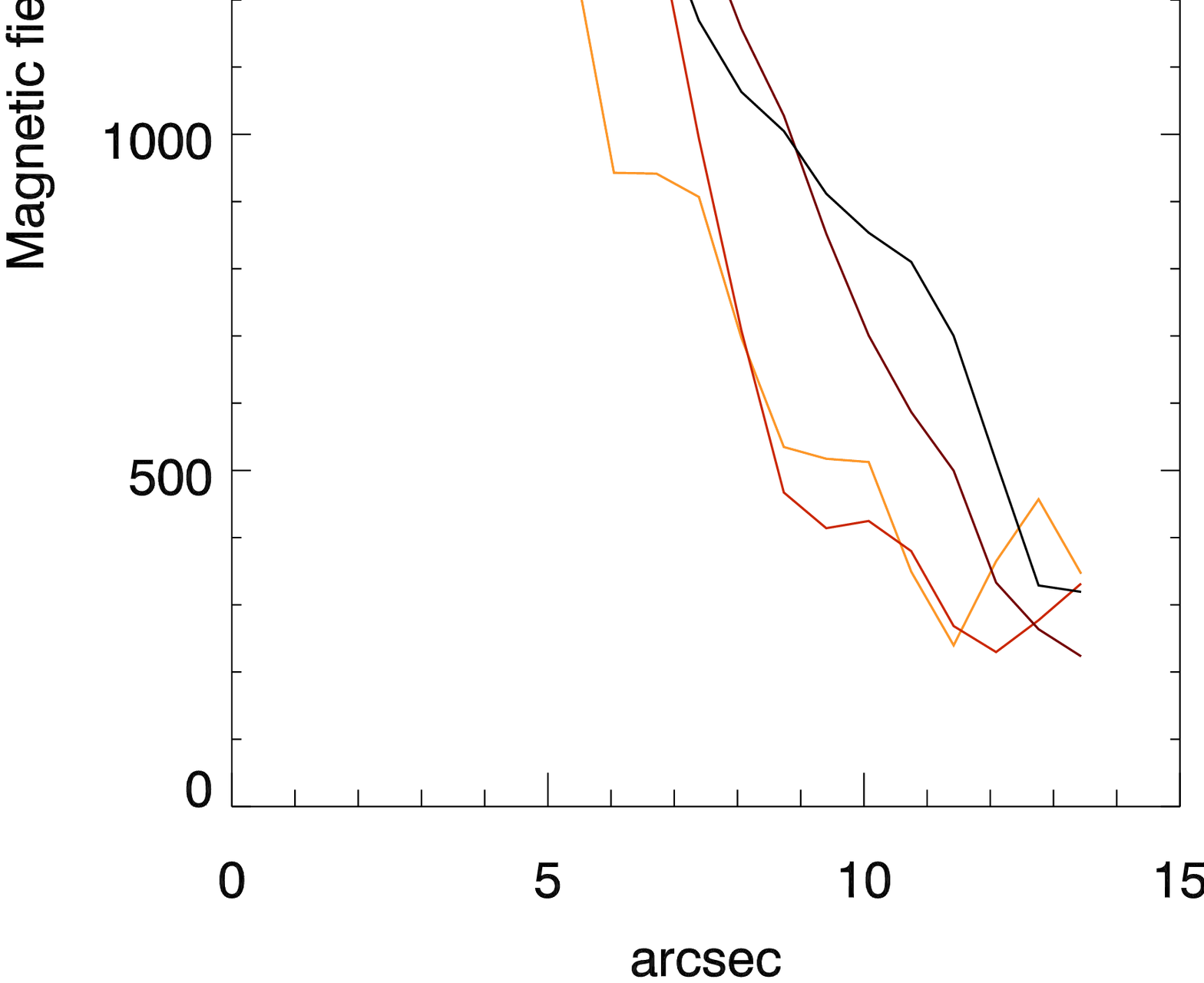}\\
\includegraphics[trim=35 10 10 30, clip, scale=0.3]{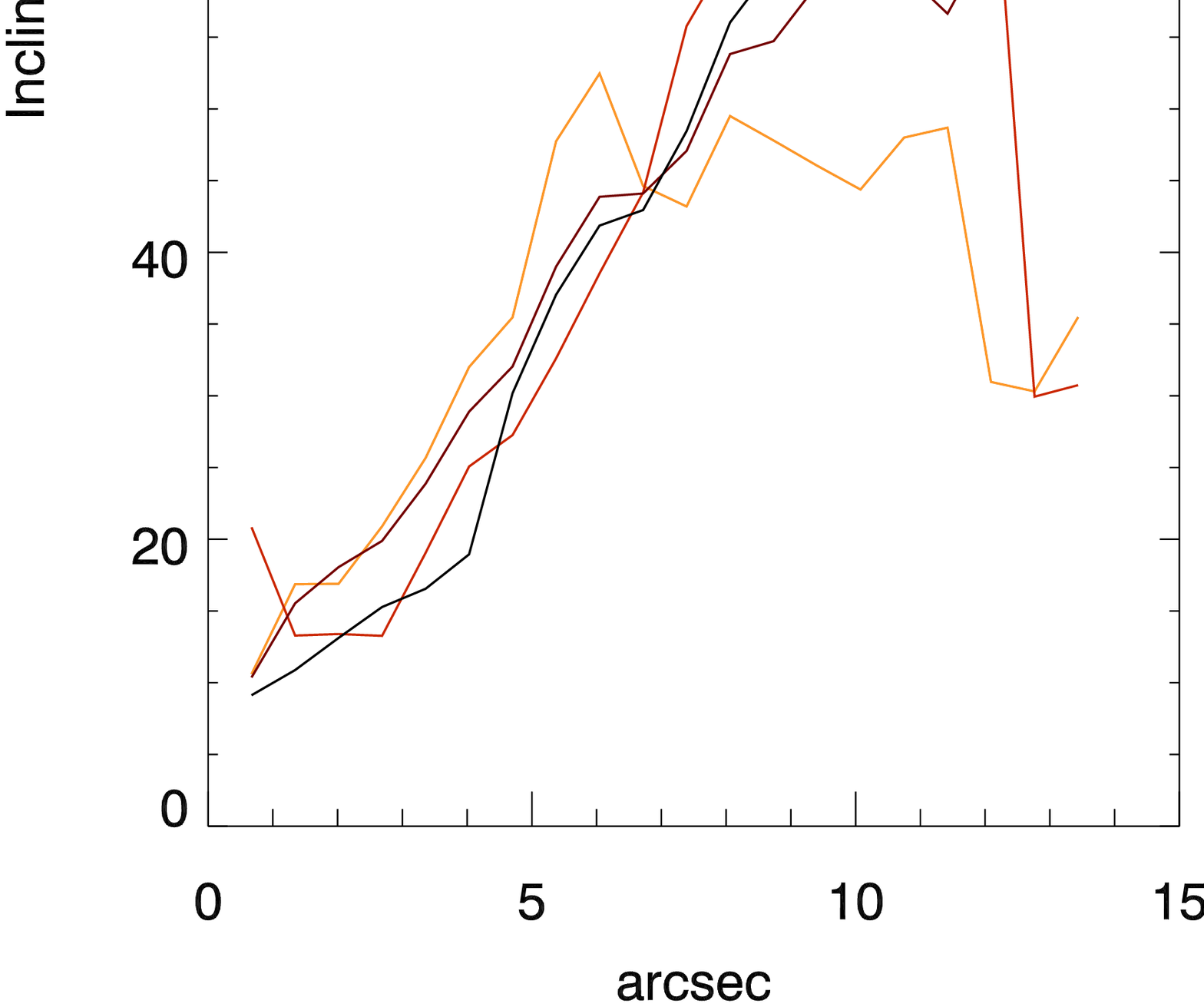}
\includegraphics[trim=35 10 10 30, clip, scale=0.3]{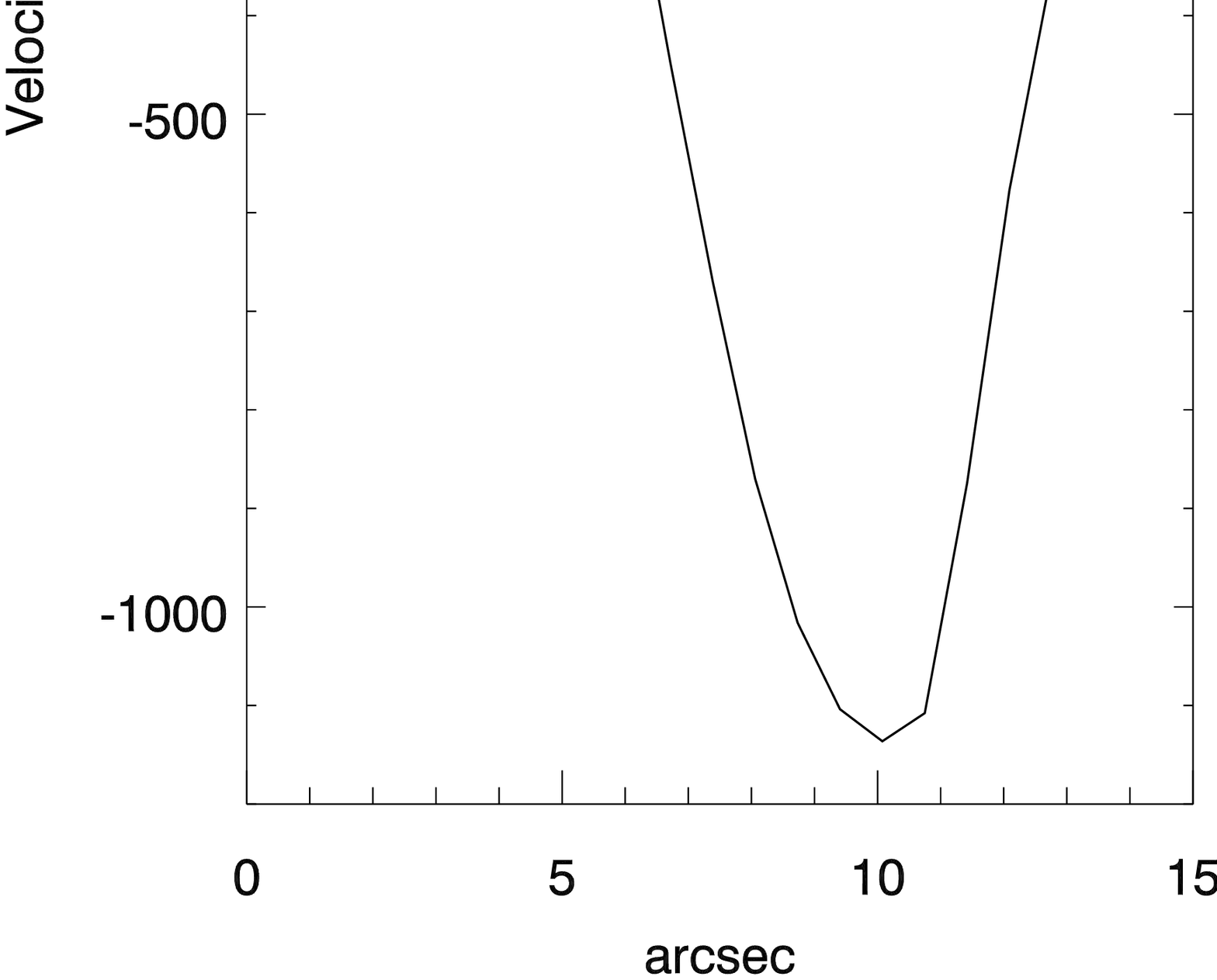}
\caption{Plots showing the variation of the continuum intensity (top left panel), the magnetic field strength (top right panel), the magnetic field inclination (bottom left panel), and the velocity along the line of sight (bottom right panel) along the region of penumbra reformation marked by a segment in the bottom right panels of Figures~\ref{fig1}, \ref{fig2}, \ref{fig3}, and~\ref{fig7}. 0 corresponds to the end of the segment towards the sunspot center. \label{fig9}} 
\end{center}
\end{figure}

The penumbra reformation seems to be accompanied by an increase of the magnetic field strength along the radial cut (see the top right panel of Figure~\ref{fig9}). This increase is almost equal along the whole segment length. We remark that this increase is limited to that area, in fact, as we already stated above, the magnetic flux of the whole sunspot decreased during the observation time interval (Figure~\ref{fig2bis}).

It is interesting also the variation of the inclination in the same area (bottom left panel of Figure~\ref{fig9}). We can see clearly that the magnetic field changes its inclination in time. On May 20 at 00:10 UT the magnetic field inclination varies from 0$^{\circ}$ to 50$^{\circ}$, while in the subsequent time interval near the outer end of the segment we see that the inclination increases up to 80$^{\circ}$, i.e., corresponding to magnetic fields bending to the photosphere. Moreover, the fact that the peak of the inclination moves outwards in time could be interpreted as a further evidence that the magnetic field restoring the penumbra may come from the upper layers of the solar atmosphere. 

During the penumbra reformation we also see a significant variation of the plasma velocity along the LOS. In fact, although at that time the sunspot was already in the Western hemisphere, looking at the disc center side of the sunspot, on May 20 at 05:58 UT positive velocity (i.e., downflow) has been found. This direction of the plasma flow confirms the reversal of the Evershed flow direction at photospheric level before and after the penumbral filament settlement. In fact, as reported in previous works \citep[see][]{Mur16}, this downflow corresponds to the counter-Evershed flow, i.e., a flux inwards to the spot center. This flow became clearly outwards after the penumbra reformation was completed.

\subsection{Chromospheric structures}

The observations taken along the \ion{Ca}{2} 854.2~nm and H$\alpha$ 656.28~nm lines by IBIS provide other interesting information about the magnetic configuration overlying the sunspot at the chromospheric level during the decay and restoring phases. This is important to ascertain whether these processes show some manifestations in the upper layers of the chromosphere.
 
\begin{figure}
   \centering
\includegraphics[trim=180 330 160 280, clip, scale=0.75]{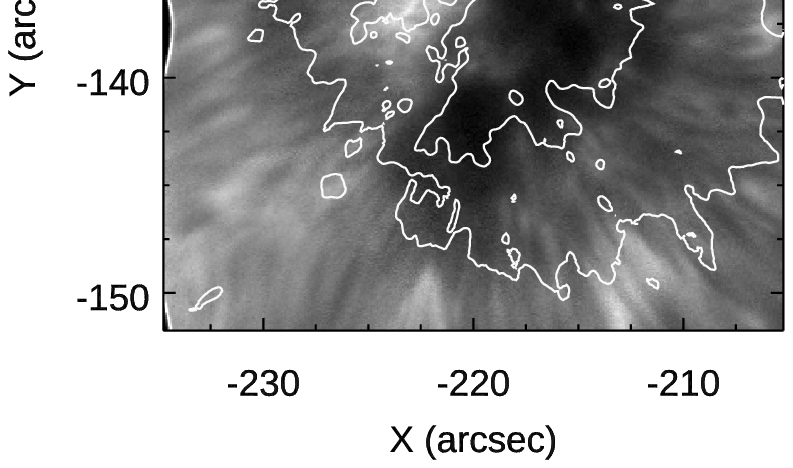}
\includegraphics[trim=205 330 50 280, clip, scale=0.75]{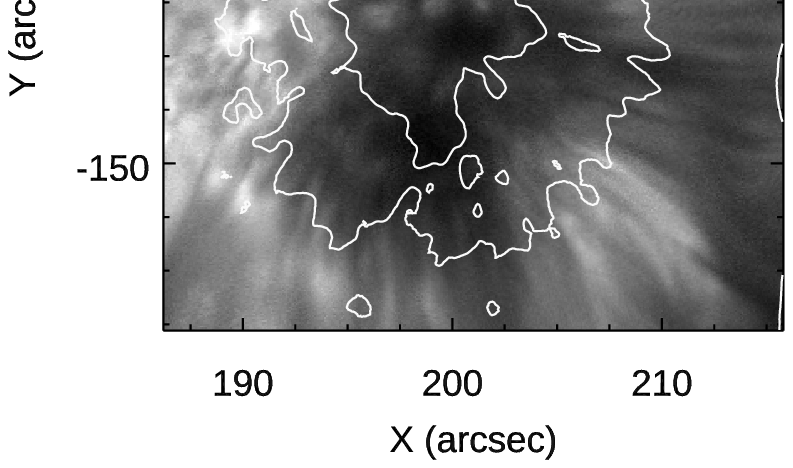}\\
\includegraphics[trim=180 300 160 280, clip, scale=0.75]{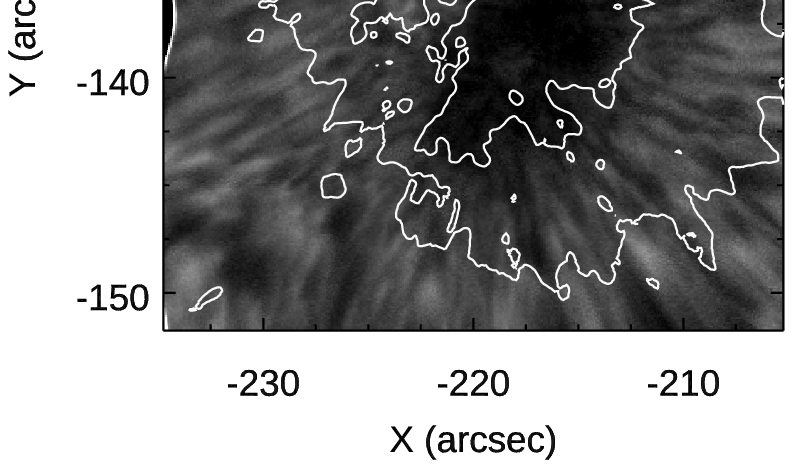}
\includegraphics[trim=205 300 50 280, clip, scale=0.75]{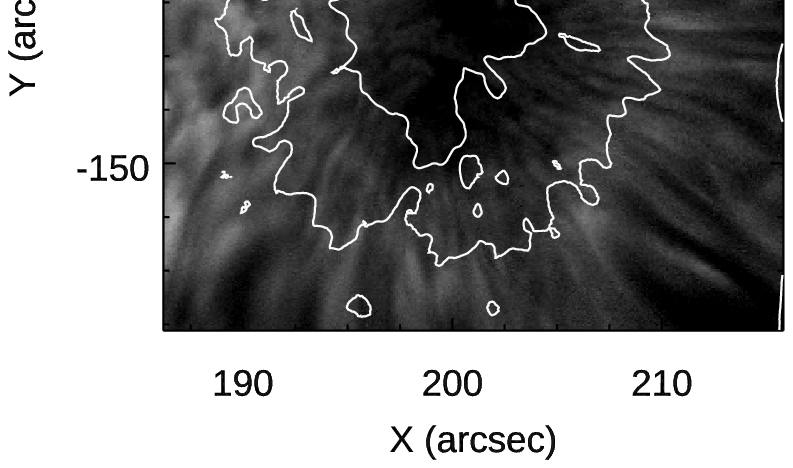}\\
 \caption{IBIS images taken on May 18 (left panels) and on May 20 (right panels) in the core of the \ion{Ca}{2} 854.2~nm line (top panels) and the H$\alpha$ 656.28~nm line (bottom panels). The inner and the outer contours indicate the umbra-penumbra and the penumbra-quiet Sun boundaries, respectively. The FOV is the same of Figure~\ref{fig5}. \label{fig8}}
\end{figure}

On May 18 in the Ca \ion{Ca}{2} we observe some brightenings in the Northern part of the IBIS FOV and in the region corresponding to the LB (top left panel of Figure~\ref{fig8}). These intensity enhancements may be attributed to the interaction between the sunspot magnetic field and the surrounding magnetic field during the fragmentation of the umbra into several parts and to their disappearance. However, at that time the sunspot appeared surrounded by radial filaments forming the so-called super penumbra \citep{Sol92}, which extends towards the quiet Sun larger than the underlying photospheric counterpart. No brightenings were visible in the H$\alpha$ line on May 18 (bottom left panel of Figure~\ref{fig8}). Conversely, the alternating bright and dark filaments around the umbra seem to be characterized by a higher contrast in comparison to those observed in \ion{Ca}{2}. It is noteworthy that in both the chromospheric lines the radial distribution of the super penumbra was still present also in the Eastern decaying portion of the sunspot. This support that disruption of the penumbra may be a bottom-up process.

On May 20, when the penumbra resettlement is still ongoing, we do not see yet the super penumbra sectors in the Eastern side of the sunspot. However, in \ion{Ca}{2} (top right panel of Figure~\ref{fig8}) we clearly see bright knots with a size of few arcseconds outside the Eastern edge of the forming sectors of the penumbra. The spatial distribution of these knots seems to be correlated with the length of the forming filaments in the Eastern side of the penumbra. Some bright knots are located about 10\arcsec{} far from the edge of the penumbra in the North-Eastern and South-Eastern sides of the sunspot. In contrast, they are located closer to the edge of the forming penumbral filaments in the Eastern side. We remark that we do not see bright knots on the other sides of the sunspot, where the penumbra remained stable. The IBIS images taken in the center of the H$\alpha$ line on May 20 (bottom right panel of Figure~\ref{fig8}) do not show these knots.

%\textcolor{red}{We plan to study the chromospheric behaviour using IBIS observations in a follow-up %paper. {Per...?}}

\section{Discussion and Conclusions}

The target of this work was suitable to shed light on the mechanisms of penumbra formation by studying the decay and subsequent restoration of a penumbral sector in the leading spot of AR NOAA 12548. We can confirm the decay of part of the sunspot by the outwards displacement of the MMFs observed in the SHARP maps of the strength and inclination of the magnetic field. Moreover, our observations are compatible with the hypothesis that the formation of the LB is also a manifestation of the decay of the magnetic field strength, as a consequence of the rearrangement of the pre-existing magnetic field of the sunspot due to its interaction with surrounding flux. Using HMI data we found that the North-Eastern part of the sunspot was also characterized by a displacement of magnetic flux towards the surrounding network. The interaction between the sunspot magnetic flux and the surrounding field is also documented by some brightenings visible in the high resolution images taken by IBIS in \ion{Ca}{2}. We can not exclude that part of the horizontal magnetic fields in the decaying penumbra became vertical, as observed by \citet{Ver18}, as a consequence of the reconnection with the neighboring magnetic field, although the presence of the super penumbra while the photospheric penumbra was not visible anymore supports the idea of the penumbra disruption likely being a bottom-up process.

It is worthy noting that the absence of local or temporal increase of the magnetic flux during the reformation of the penumbra points to a rearrangement of the magnetic field configuration occurring during the process. Therefore, the penumbra restoration appears to be triggered by a magnetic field becoming more inclined, which is able to activate the penumbral magneto-convective mode \citep[e.g.,][]{Jur14,Jur17b}.

The restoring phase of the penumbra seems to support the scenario proposed by \citet{Rom13, Rom14, Mur16}, i.e., the penumbra forms by bending of the magnetic field lines already present in the higher layers of the solar atmosphere. In fact, after the disappearance of the penumbral sector we did not observe new magnetic flux emergence. On the contrary, we found that the area where the penumbral sector of the sunspot disappeared was characterized by a magnetic field whose inclination varied progressively from vertical to horizontal (compare the bottom panels of Figure~\ref{fig3}). This is clearly visible in the accompanying online movie ($Inclination\_AR12348.wmv$) of the inclination maps obtained by HMI data. We found that new filamentary structures started to appear at the edge between the umbra and the quiet Sun and filled progressively the gap of the penumbra sector with an uncombed magnetic field, characterized by the presence of more and more elongated filamentary structures with an inclination of about 90$^{\circ}$. This evolution can be interpreted as the manifestation of the magnetic field lines of the magnetic canopy that change their inclination till they touch the photosphere. 
After the penumbra was completely reformed, evidence for the removal of the azimuth angle discontinuity, which was cospatial to the LB during penumbral decay, strongly supports that the restoration process led to a new stable configuration of the sunspot.

The presence of bright knots detected by IBIS in the Ca II line at chromospheric level, in the side of the sunspot where the penumbra was restoring, could be the signature of this bending of the magnetic field lines. In fact, we can explain these brightenings as due to the interaction between the tails of the forming inter-spines and the lower atmosphere. We also observed in HMI data some patches characterized by a magnetic field polarity opposite to that of the sunspot only in the latest phase of the penumbra restoration, i.e., when the tails of the penumbral filaments reached the photosphere and the complete penumbra appeared again in the continuum images.

The high-resolution spectro-polarimetric data taken with IBIS and inverted with SIR confirm this scenario. In fact, on May 20 at about 13:30~UT, when the restoring process had started since some hours and was still ongoing, at photospheric level we observe the presence of a vertical magnetic field in the sector where the penumbra was restoring (bottom right panel of Figure~\ref{fig5}). This result reinforces the hypothesis that the penumbra forms from an initial vertical field which changes its inclination. Unfortunately, due to the worsening of the seeing conditions we did not take other IBIS data during the subsequent hours of the day, therefore we could not follow the variation of the magnetic field inclination exploiting the high resolution of the instrument.

Another interesting evidence of the settlement of the horizontal penumbral filaments inside the restoring sector of the penumbra was the transition from the counter-Evershed flow to the classical one. This transition, already observed in \citet{Mur16, Mur18}, confirms that the development of the magnetic field configuration, typical of the penumbra, takes several hours and that the classical Evershed flow starts only when the sinking magnetic field dips below the solar surface.

We think that future high resolution observations carried out by the next generation telescopes, like the D.~K. Inouye Solar Telescope \citep{Kei10} and the European Solar Telescope \citep{Col10}, or the future use of IBIS at the focus of a large aperture solar telescope may provide further interesting data to strengthen the results described in this work.

%\begin{figure}
%\begin{center}
%\includegraphics[trim=70 180 10 230, clip, scale=0.4]{subFOV_doppler_plot_calib1.eps}
%\includegraphics[trim=70 180 10 230, clip, scale=0.4]{subFOV_doppler_plot_calib2.eps}\\
%\caption{\bf . \label{fig12}} 
%\end{center}
%\end{figure}

\acknowledgments

The authors wish to thank the DST staff for its support during the observing campaigns. The research leading to these results has received funding from the European Union’s Horizon 2020 research and innovation programme under grant agreement no. 739500 (PRE-EST project) and no. 824135 (SOLARNET project).
This work was supported by the Italian MIUR-PRIN 2017 on Space Weather: impact on circumterrestrial environment of solar activity, by Space Weather Italian COmmunity (SWICO) Research Program, and by the Universit\`{a} degli Studi di Catania (Piano per la Ricerca Universit\`{a} di Catania 2016-2018 – Linea di intervento 1 \textquotedblleft Chance\textquotedblright; Linea di intervento 2 \textquotedblleft Ricerca di Ateneo - Piano per la Ricerca 2016/2018\textquotedblright; \textquotedblleft Fondi di ateneo 2020-2022, Universit\`{a} di Catania, linea Open Access\textquotedblright).

%% To help institutions obtain information on the effectiveness of their
%% telescopes, the AAS Journals has created a group of keywords for telescope
%% facilities. A common set of keywords will make these types of searches
%% significantly easier and more accurate. In addition, they will also be
%% useful in linking papers together which utilize the same telescopes
%% within the framework of the National Virtual Observatory.
%% See the AASTeX Web site at http://www.journals.uchicago.edu/AAS/AASTeX
%% for information on obtaining the facility keywords.

%% After the acknowledgments section, use the following syntax and the
%% \facility{} macro to list the keywords of facilities used in the research
%% for the paper.  Each keyword will be checked against the master list during
%% copy editing.  Individual instruments or configurations can be provided 
%% in parentheses, after the keyword, but they will not be verified.

{\it Facilities:} \facility{DST (IBIS), SDO (AIA, HMI)}.

%% Appendix material should be preceded with a single \appendix command.
%% There should be a \section command for each appendix. Mark appendix
%% subsections with the same markup you use in the main body of the paper.

%% Each Appendix (indicated with \section) will be lettered A, B, C, etc.
%% The equation counter will reset when it encounters the \appendix
%% command and will number appendix equations (A1), (A2), etc.

%\appendix

%\section{Appendix material}

\clearpage

\end{document}